\documentclass{aastex631}
\usepackage{amsmath}
\usepackage{booktabs}

\usepackage{chngpage}
\usepackage{multirow}
\usepackage{subfigure}
\usepackage{threeparttable} 
\usepackage{float}
\usepackage{hyperref}
\usepackage[section]{placeins} 
\usepackage[figuresright]{rotating}

\usepackage{color}

\shorttitle{The Energy Budget in the Jet of High-frequency Peaked BL Lacertae Objects}
\shortauthors{Zhao et al.}

\begin{document}

\title{The Energy Budget in the Jet of High-frequency Peaked BL Lacertae Objects }

\correspondingauthor{Y. G. Zheng }
\email{$^\ast$ynzyg@ynu.edu.cn}
\email{$^\dagger$kangshiju@alumni.hust.edu.cn}

\author[0009-0008-3339-2035]{X. Z. Zhao}
\affiliation{Department of Physics, Yunnan Normal University, Kunming, Yunnan, 650092, People's Republic of China}

\author{H. Y. Yang}
\affiliation{College of Physics and Information Engineering, Zhaotong University, Zhaotong, 657000, People's Republic of China}

\author[0000-0003-0170-9065]{Y. G. Zheng$^\ast$}
\affiliation{Department of Physics, Yunnan Normal University, Kunming, Yunnan, 650092, People's Republic of China}

\author[0000-0002-9071-5469]{S. J. Kang$^\dagger$}
\affiliation{School of Physics and Electrical Engineering, Liupanshui Normal University, Liupanshui, Guizhou, 553004, People's Republic of China}

\begin{abstract}

Energy equipartition and the energy budget in the jet are import issues for the radiation mechanism of blazars.
Early work predominantly concentrated on flat-spectrum radio quasars and a limited number of BL Lacertae objects (BL Lacs).
In this paper, we compile 348 high-frequency peaked BL Lac objects (HBLs) based on the catalog of active galactic nuclei (4LAC-DR3) from Fermi-LAT, and employ \textit{JetSet} to fit the spectral energy distributions (SEDs) of these HBLs in the framework of the one-zone lepton model.
We aim to determine whether the energy budget is reasonable and whether the energy equipartition  is satisfied in HBLs.
The results of the statistical analysis suggest that:
(1) SEDs of HBLs can be reproduced well by using the one-zone lepton model; however it cannot achieve the energy equalization, and the relativistic electron energy density is far greater than the magnetic field energy density, $U_{e} \gtrsim100 U_{B}$;
(2) the majority of the HBLs are located in the $t_{cool}$$<$$t_{dyn}$ region (where the horizontal coordinate represents the jet power of electrons, while the ordinate indicates the ratio between the dynamic time scale to the cooling timescale), and the jet kinetic power of HBLs is greater than the jet power of radiation; there is a very low radiation efficiency, we deduce that HBLs may have optically thin advection-dominated accretion flows; 
(3) the $\log\epsilon_{B}$ of HBLs is less than zero, which indicates that the jet kinetic power of HBLs is not affected by Poynting flux;
(4) the relationships with $U_{e} >U_{Syn}\sim U_{B}$, $L_{e}\sim L_{p}>L_{B}\sim L_{rad}$ and $\log\epsilon_{e}>0.5$ are established.
These relations indicate that most of the energy of HBLs is stored in the population of low-energy electrons.

\end{abstract}


\section{Introduction} \label{sec:intro}

Active galactic nuclei (AGNs) are some of the most mumerous extragalactic objects in astronomy, among which blazars are an extreme subclass of radio-loud AGNs that exhibit extreme properties such as rapid variability, high luminosity, and high polarization (\citealt{1992ApJ...398..454W,1998A&AS..132...83B,2010MNRAS.402..497G,2014ApJS..215....5K,2016ApJS..226...20F,2022ApJS..262...18Y}).
The equivalent width (EW) of optical emission lines divides blazars into flat-spectrum radio quasars (FSRQs) and BL Lacertae (BL Lac) objects , where FSRQs have strong broad emission lines (EW $\geqslant5\mathring{\mathrm{A}}$) and BL Lac objects have weak or no broad emission lines (EW $<5\mathring{\mathrm{A}}$) (\citealt{1995PASP..107..803U}).
Although the two subclasses exhibit numerous observational similarities, disparities in the characteristics of broad emission line imply distinct physical mechanisms underlying the production of these lines.
The multiwavelength spectral energy distributions (SEDs) of the blazars show a typical double-hump structure (\citealt{1996ApJ...463..555I,2010ApJ...716...30A,2010MNRAS.402..497G,2014ApJS..215....5K,2019ApJ...873....7Z,2020ApJS..247...49X,2020MNRAS.499.1188Z,2021ApJ...915...59Z,2024ApJ...962...22Z,2021ApJ...916...93Z,2024MNRAS.528.7587H}).
The low-energy hump extends from the radio to the X-ray band, and its origin is effectively accounted for by synchrotron (Syn) mechanisms. On the other hand, the high-energy hump is in the MeV-TeV energy range. However, the mechanism that produces the high-energy hump is an outstanding issue.
The causes of this hump could be  inverse Compton scattering of relativistic electrons, either on synchrotron photons (Synchrotron Self-Compton, SSC; \citealt{1992ApJ...397L...5M,2010MNRAS.401.1570T,2018MNRAS.478.3855Z}) or on other photon groups (external Compton, EC; \citealt{1993ApJ...416..458D,1994ApJ...421..153S,2014ApJS..215....5K}). The position of the first peak in SEDs, $\nu_{p}^{S}$ (synchrotron peak frequency), classifies the sources as low-synchrotron-peaked (LSP; e.g., $\nu_{p}^{S}$ $<$ $10^{14}$ Hz), intermediate-synchrotron-peaked (ISP; e.g., $10^{14}$ Hz $<$$\nu_{p}^{S}$ $<$ $10^{15}$ Hz), and high-synchrotron-peaked (HSP; e.g., $\nu_{p}^{S}$ $>$ $10^{15}$ Hz) blazars (e.g., \citealt{1995ApJ...444..567P,2010ApJ...716...30A}).
There are noticeable obvious differences in the frequency of low-energy peaks in the SEDs distribution of BL Lac objects. \cite{2010ApJ...716...30A} also divided BL Lac objects into low-frequency peaked BL Lacs (LBLs; e.g., $\nu_{p}^{S}$ $<$ $10^{14}$ Hz), intermediate-frequency peaked BL Lacs (IBLs; e.g., $10^{14}$ Hz $<$$\nu_{p}^{S}$ $<$ $10^{15}$ Hz), and high-frequency peaked BL Lacs (HBLs; e.g., $\nu_{p}^{S}$ $>$ $10^{15}$ Hz).
Many authors (\citealt{2010MNRAS.401.1570T,2012ApJ...752..157Z,2023ApJS..268...23F}) have found that the SEDs of BL Lac objects appears to accord with a pure SSC model, especially for HBLs.

The formation mechanism of relativistic jet in AGNs is still a mystery, and many proposed models discuss jet formation.
At present, the two most mature theories are the Blandford-Znajek mechanism (\citealt{1977MNRAS.179..433B}), where the jet extracts the rotational energy of the black hole (BH), and the Blandford-Payne mechanism (\citealt{1982MNRAS.199..883B}), where the jet mainly extracts the rotational energy of the accretion disk.
In both cases, the magnetic field is import in directing power from the BH  or disk into the jet (\citealt{2003ApJ...593..667M}). 

The Poynting flux initially dominates the energy budget of the jet, which gradually transforms into the plasma's kinetic energy as the flow accelerates.
Theoretical studies of the energy dissipation process suggest that in this case the emitting electrons and the electromagnetic field may carry the same amount of energy (\citealt{2015MNRAS.450..183S}); that is, the magnetic energy flux of the large-scale jet should still account for a large part (about half) of the total jet power.
Ideally, the magnetic flux carried by the jet should be comparable to that supported by the inner accretion disk and should regulate the energy extracted from the BH (\citealt{2016MNRAS.456.2374T}).
However, from an observational point of view, the situation is unclear.
Therefore, understanding the magnitude of the magnetic field inside the jet is crucial to understanding its formation and energy budget.
By using the observed frequency dependence \citep{1998A&A...330...79L} of the position of the optically thick jet core , \cite{2014Natur.510..126Z} obtained that the jet magnetic flux at the parsec scale is related to the power of the corresponding accretion current, and is similar to the value predicted by magnetohydrodynamics.
Another reliable method for estimating the associated magnetic field in the innermost emission region of the jet is by modeling the SEDs of blazars (\citealt{1998MNRAS.301..451G,2010MNRAS.402..497G,1998ApJ...509..608T,2010MNRAS.405L..94T}).
The SEDs modeling of FSRQs by \cite{2013ApJ...768...54B} and \cite{2020ApJS..248...27T} shows that the parameters of SEDs are close to the equipartition condition between the magnetic field and the relativistic electrons.

As previously mentioned, most of the sources investigated for their magnetic field effects are FSRQs.
It is important to note that BL Lac objects and FSRQs exhibit distinct characteristics, which implies potential differences in the formation and structure of their jets. \cite{2010MNRAS.401.1570T} used the  one-zone lepton model (Syn+SSC) to model 45 BL Lac objects. Based on their results, \cite{2016MNRAS.456.2374T} studied the relativistic electron energy density and magnetic field energy density and found that the relativistic electron energy density of BL Lac objects is two orders of magnitude larger their magnetic field energy density and has lower radiation efficiency.
Since the successful launch of the Fermi Space Telescope, many AGNs have been detected in high-energy gamma rays (\citealt{2009ApJS..183...46A,2010ApJ...716...30A,2010ApJS..188..405A,2015ApJS..218...23A,2020ApJS..247...33A,2022ApJS..260...53A}), in particular, we can expand our collection of blazar samples.
Compared to previous studies, we emphasize objects such as HBLs and employ the one-zone lepton model to elucidate the observations.
In this paper, we have expanded our sample size, making it perhaps the most comprehensive collection of HBL samples modeled by incorporating a physical SED model.
The primary objective of our study is to investigate whether the energy budget is reasonable and whether HBLs achieve energy equipartition within the framework of the one-zone lepton model, while we also examine their  fundamental characteristics.
In Section~\ref{sec:sample}, we introduce our samples.
In Section~\ref{sec:model}, we present the fitting tools and the fitting process.
Section~\ref{sec:results} describes the results and provides a discussion.
Finally, we present the conclusion of this work in Section~\ref{sec:conclusion}. 
Throughout the paper, we assume the Hubble constant to be $H_{0} = 67.8$ km s$^{-1}$ Mpc$^{-1}$, the matter energy density to be $\Omega_{\rm M} = 0.307$, the radiation energy density to be $\Omega_{\rm r} = 0$, and the dimensionless cosmological constant to be $\Omega_{\Lambda} = 0.69$.

\section{The Sample}     \label{sec:sample}
\subsection{HBL Sample}

4FGL-DR3\footnote{\scriptsize{\url{https://fermi.gsfc.nasa.gov/ssc/data/access/lat/12yr_catalog/}}} (\citealt{2022ApJS..260...53A}) is the fourth comprehensive catalog of Fermi Large Area Telescope sources, drawing from 12 years of survey data within the 50 MeV-1 TeV energy range. 4LAC-DR3\footnote{\scriptsize{\url{https://fermi.gsfc.nasa.gov/ssc/data/access/lat/4LACDR3/}}} (\citealt{2022ApJS..263...24A}) is an AGN catalog that is derived from the 4FGL-DR3 and encompasses 3743 blazars, comprising 1458 BL Lac objects, 792 FSRQs, and 1493 blazars of uncertain type.
Our compilation includes 425 HBLs identified in the 4LAC-DR3 (High Latitude Sources) catalog, with 292 possessing usable redshifts.
For sources lacking redshift information, we initially search SIMBAD\footnote{\scriptsize{\url{http://simbad.cds.unistra.fr/simbad/}}} (\citealt{2000A&AS..143....9W}), identifying the redshifts of 61 HBLs. Subsequently, we utilize our sample's average redshift ($\langle z_{}\rangle\simeq0.399$) to estimate redshifits for the remaining sources.

\subsection{Multiband Data}

The multiband data for 404 of the sample of 425 HBLs we constructed were derived from the BlaST (acronym for blazar synchrotron tool or blazar SED tool)\footnote{\scriptsize{\url{https://github.com/tkerscher/blast/blob/master/4LAC.zip}}} (\citealt{2022A&C....4100646G}), which is a subsample of the master list of blazars, selected on the basis of the availability of sufficient multifrequency data and to ensure that all blazars types (LBLs, IBLs, and HBLs) and data combinations (jet emission plus other non-jet-related components) are adequately represented. 
The SEDs of each blazar in the BlaST sample was assembled using the VOU-Blazars (V1.94) tool (\citealt{2020A&C....3000350C}), which retrieves multiband data from 71 catalogs and spectral databases from different online services using Virtual Observatory\footnote{\scriptsize{\url{https://ivoa.net/}}} protocols.
Once the data have been downloaded, VOU-Blazars automatically converts them to homogeneous SED units, and then the optical measurements are de-reddening and converted to soft X-ray measurements to remove the effects of Galactic absorption. 
The remaining 21 HBLs were not present in the BlaST sample, and we have  collected multiband data for these 21 HBLs from 2008 October to 2023 October, using data from the Space Science Data Center (SSDC) SED Builder, an online service developed by SSDC\footnote{\scriptsize{\url{https://tools.ssdc.asi.it}}} ( \citealt{2011arXiv1103.0749S}). The SSDC synthesizes flux data across the radio to $\gamma$-ray (including TeV) bands, from various catalogs archival data, etc. 

Our collection of SEDs contains data from different time periods, which is unavoidable since measurements are rarely made simultaneously. Hence, the result predicted by the \textit{JetSet} is a time-averaged one, as typically used in the literature \citep{2022A&C....4100646G}.

Then, we fit the SEDs of these sources and filter them according to our fitting results.
First, we remove sources whose peak frequency of synchrotron radiation is less than $10^{15}$ Hz, and secondly we exclude the sources with poor fitting results ($\chi^{2}/dof \geq 30$).
In the end, we identify 348 HBLs, 298 of which have reliable redshifts. Table~\ref{tab:1} presents information about these sources.

\begin{sidewaystable*}[!htp]
\tablenum{1}
	\centering
	\small
		\begin{center}
			\caption{The Parameters Used to Fit the SEDs }    
			\label{tab:1}
			\setcounter{table}{1}
			\renewcommand{\thetable}{1/arabic{table}}
			\renewcommand\arraystretch{2}
			\begin{adjustwidth}{-3.6cm}{1cm}
				\scalebox{1}{
					\begin{tabular}{ccccccccccccccc}
						\hline\hline	 			
						$\rm 4FGL\;Name$&$\rm Association$  & ${\rm z}$ & $ log v_{p}^{S}$ & $log v_{p}^{SSC}$ & $p_1$ & $p_2$ & {$log N_0$} & {$log \delta$}  & {$log B$}  & {$log R$}  & {$log \gamma_{min}$}  & {$log \gamma_{break}$}  & {$log \gamma_{max}$} & {$\chi^{2}/dof$}\\
						\normalsize(1) & \normalsize(2) & \normalsize(3) &\normalsize(4) & \normalsize(5)   &\normalsize(6) &\normalsize(7) &\normalsize(8) &\normalsize(9) &\normalsize(10) &\normalsize(11) &\normalsize(12) &\normalsize(13) &\normalsize(14) &\normalsize(15)   \\
						\hline
						\centering
  {$ \rm J0013.9-1854$} & {$ \rm RBS\;0030$} & 0.09  & 16.37  & 24.20  & 2.33  & 3.50  & 2.47  & 1.36  & 0.14  & 15.05  & 1.95  & 4.15  & 5.92  & 10.06  \\
{$ \rm J0014.1-5022$} & {$ \rm RBS\;0032$} & 0.01  & 17.02  & 24.38  & 2.33  & 3.18  & 3.77  & 1.12  & 0.22  & 13.91  & 2.16  & 4.43  & 5.30  & 4.53  \\
{$ \rm J0022.0+0006$} & {$ \rm RX\;J0022.0+0006$} & 0.31  & 16.63  & 24.24  & 2.40  & 3.65  & 3.14  & 1.61  & 0.28  & 14.49  & 1.72  & 4.13  & 5.54  & 7.74  \\
{$ \rm J0033.5-1921$} & {$ \rm KUV\;00311-1938$} & 0.61  & 15.55  & 24.92  & 2.11  & 3.81  & 1.59  & 1.60  & -2.15  & 17.45  & 0.53  & 4.86  & 6.69  & 5.33  \\
{$ \rm J0043.7-1116$} & {$ \rm 1RXS\;J004349.3-111612$} & 0.26  & 15.36  & 24.35  & 2.06  & 3.56  & 1.31  & 1.70  & -1.63  & 16.22  & 1.62  & 4.35  & 6.10  & 2.79  \\
{$ \rm J0045.3+2128$} & {$ \rm GB6\;J0045+2127$} & {{$2.07^{b}$}} & 15.83  & 24.96  & 2.13  & 4.00  & 0.77  & 1.69  & -1.72  & 17.45  & 1.25  & 4.89  & 6.66  & 6.16  \\
{$ \rm J0051.2-6242$} & {$ \rm 1RXS\;J005117.7-624154$} & 0.30  & 15.40  & 24.64  & 1.63  & 3.50  & -0.69  & 0.77  & -1.82  & 18.14  & 0.49  & 4.86  & 6.27  & 2.89  \\
{$ \rm J0134.5+2637^{c} $} & {$ \rm RX\;J0134.4+2638$} & {$0.57^{b}$} & 15.98 & 22.43 & 2.86  & 4.63  & 1.99  & 1.39  & -0.65 & 16.2  & 2.4   & 4.74  & 6.74  & 10.86 \\
{$ \rm J0136.5+3906$} & {$ \rm B3\;0133+388$} & {\nodata}  & 16.16  & 24.64  & 1.37  & 2.99  & -2.39  & 0.93  & -1.87  & 18.58  & 1.84  & 4.60  & 6.31  & 2.12  \\
{$ \rm J0352.0-3702^a$} & {$ \rm 1E\;0350.0-3712$} & 0.17 & 15.69  & 25.75  & 1.00  & 3.00  & -2.64  & 1.70  & -2.94  & 17.20  & 3.33  & 4.71  & 6.71  & 13.93  \\					 
					  								
						\hline
				\end{tabular}}\\
			\end{adjustwidth}
		\end{center}  
			\tablecomments{Columns (1): the 4FGL name of sources;columns (2): the associated source name; columns (3): the redshift; columns (4): the synchrotron peak frequency ($\mathrm{Hz}$); columns (5): the synchrotron self-Compton peak frequency($\mathrm{Hz}$); columns (6)-(7): the electron spectral index; columns (8): the normalization constant ($cm^{-3}$); columns (9): the Doppler factor; columns (10): the magnetic field ($G$); columns (11): the the radiation region ($cm$); columns (12): the minimum Lorentz factor; columns (13): the broken Lorentz factor; columns: is the maximum Lorentz factor;  columns (15): the $\chi^{2}/dof$ values.
				 (This table is available in its entirety in machine-readable form.  }		
		\begin{tablenotes}   
			\footnotesize               
			\item[1]a.The multiband data from SSDC.
			\item[2]b.Redshift from SIMBAD.
			\item[3]c.Add functional dependences in the third step of the fitting process.        			
		\end{tablenotes}           		
\end{sidewaystable*}

\begin{deluxetable*}{cc}
	\tablenum{2}
		\tablecaption{The fit range and the default Values of the
			parameters
			\label{tab:2}}
			\tablewidth{0pt}
			\tablehead{
				\colhead{Wave Band } &  \colhead{Log Range (Hz)}
			}
			\decimalcolnumbers
			\startdata
			Radio & [6,~10] \\
			$Radio\_mm$ &[10,~11] \\
			$mm\_IR$ &[11,~13] \\
			$IR\_Opt$ &[13,~14] \\
			$Opt\_UV$ &[14,~16] \\
			$UV\_X$ &[15,~17.5] \\
			BBB &[15,~16] \\
			X &[16,~19] \\
			Fermi &[22.38,~25.38] \\
			TeV &[25,~28.38] \\
			\enddata
	\begin{adjustwidth}{1cm}{0cm}
		\tablecomments{mm: millimeter, IR: infrared, Opt: optics, UV: ultraviolet, BBB: big blue bump. These values come from \textit{JetSet}.}	
	\end{adjustwidth}
\end{deluxetable*}

\section{Multiband SED fitting based on \textit{JetSet}}  \label{sec:model}

\subsection{The \textit{JetSet} fitting tool}

\textit{JetSet}\footnote{\scriptsize{\url{https://jetset.readthedocs.io/en/1.2.2/index.html}}} (\citealt{2009A&A...501..879T,2011ApJ...739...66T,2020ascl.soft09001T}) is an open-source C/Python framework to reproduce radiative and accelerative processes acting in relativistic jet, and galactic objects (beamed and unbeamed), allowing to fit the numerical models to observed data. The main features of this framework are:

$\bullet$ Handling observed data involves rebinning, defining data sets, connecting to $astropy$\footnote{\scriptsize{\url{https://www.astropy.org/}}} \citep{2013A&A...558A..33A,2018AJ....156..123A,2022ApJ...935..167A} tables, and defining quantities for complex numerical radiative scenarios: SSC, EC, and EC against the Cosmic Microwave Background (CMB).

$\bullet$ During the prefitting stage, the model is constrained by leveraging precise and previously published phenomenological trends.
The process commences with well-established parameters, such as spectral indices, peak fluxes, frequencies, and spectral curvatures, which the code automatically evaluates.
The prefitting algorithm utilizes these parameters to generate an initial model in alignment with the phenomenological trends implemented in \textit{JetSet}.
The subsequent fitting of multiwavelength SEDs follows these approaches, the frequentist approach (iminuit\footnote{\scriptsize{\url{https://iminuit.readthedocs.io/en/stable/index.html}}} (V2.22.0 \citep{dembinski_2023_8070217}) and the Bayesian Markov Chain Monte Carlo (MCMC) sampling (emcee\footnote{\scriptsize{\url{https://emcee.readthedocs.io/en/stable/index.html}}} \citep{2013PASP..125..306F}).

$\bullet$ Self-consistent temporal evolution of the plasma under radiative and accelerative processes, and adiabatic expansion.
Implementation of both first-order and second-order (stochastic acceleration) processes.

\begin{figure*}[!htbp]
	\centering
	\begin{adjustwidth}{-0.5cm}{-1.5cm}
		
		\subfigbottomskip=2pt
		\subfigcapskip=-2pt
		\subfigure{
			\includegraphics[width=0.45\linewidth]{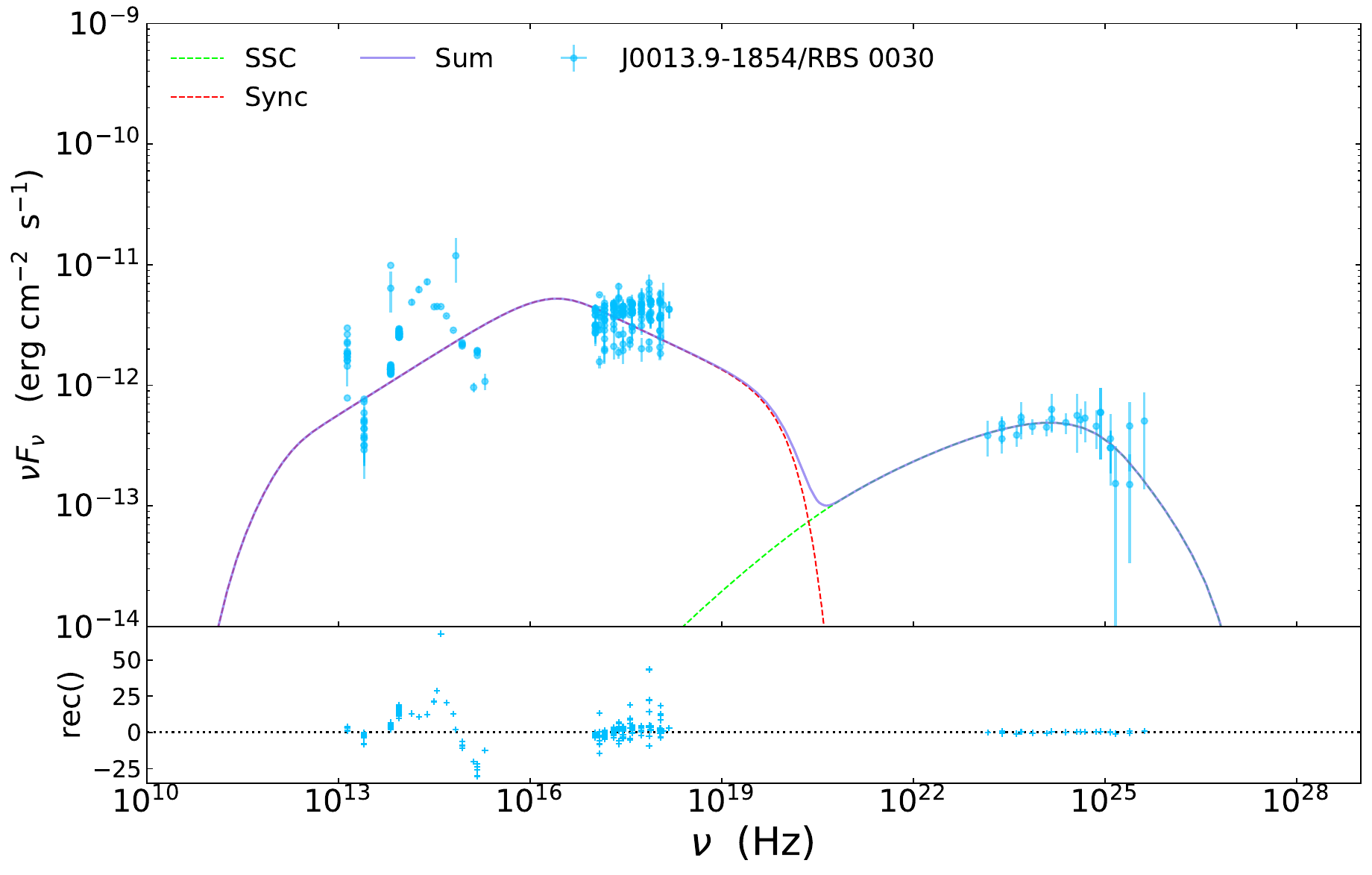}}
		\subfigure{
			\includegraphics[width=0.45\linewidth]{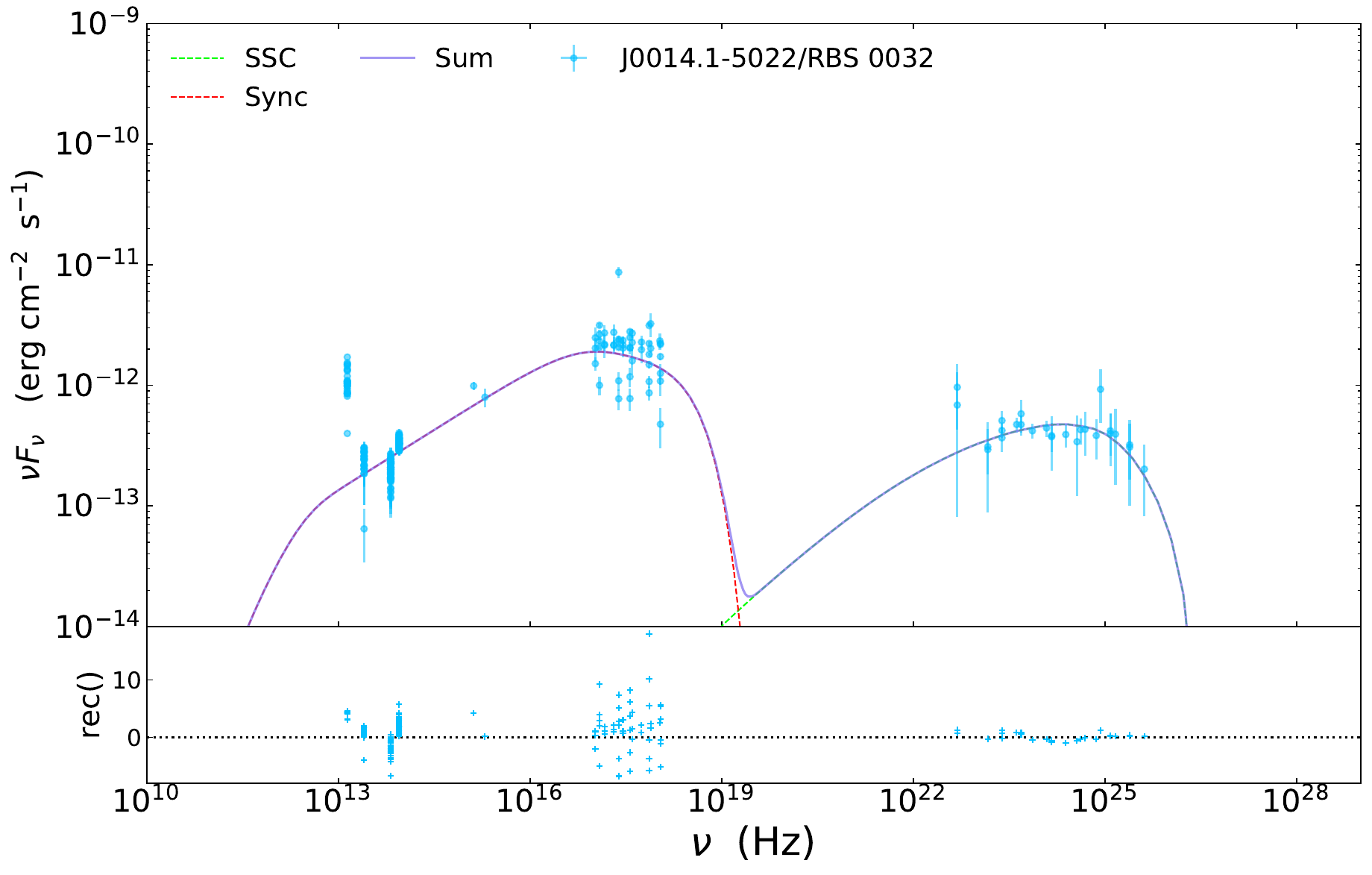}}
		
		\subfigure{
			\includegraphics[width=0.45\linewidth]{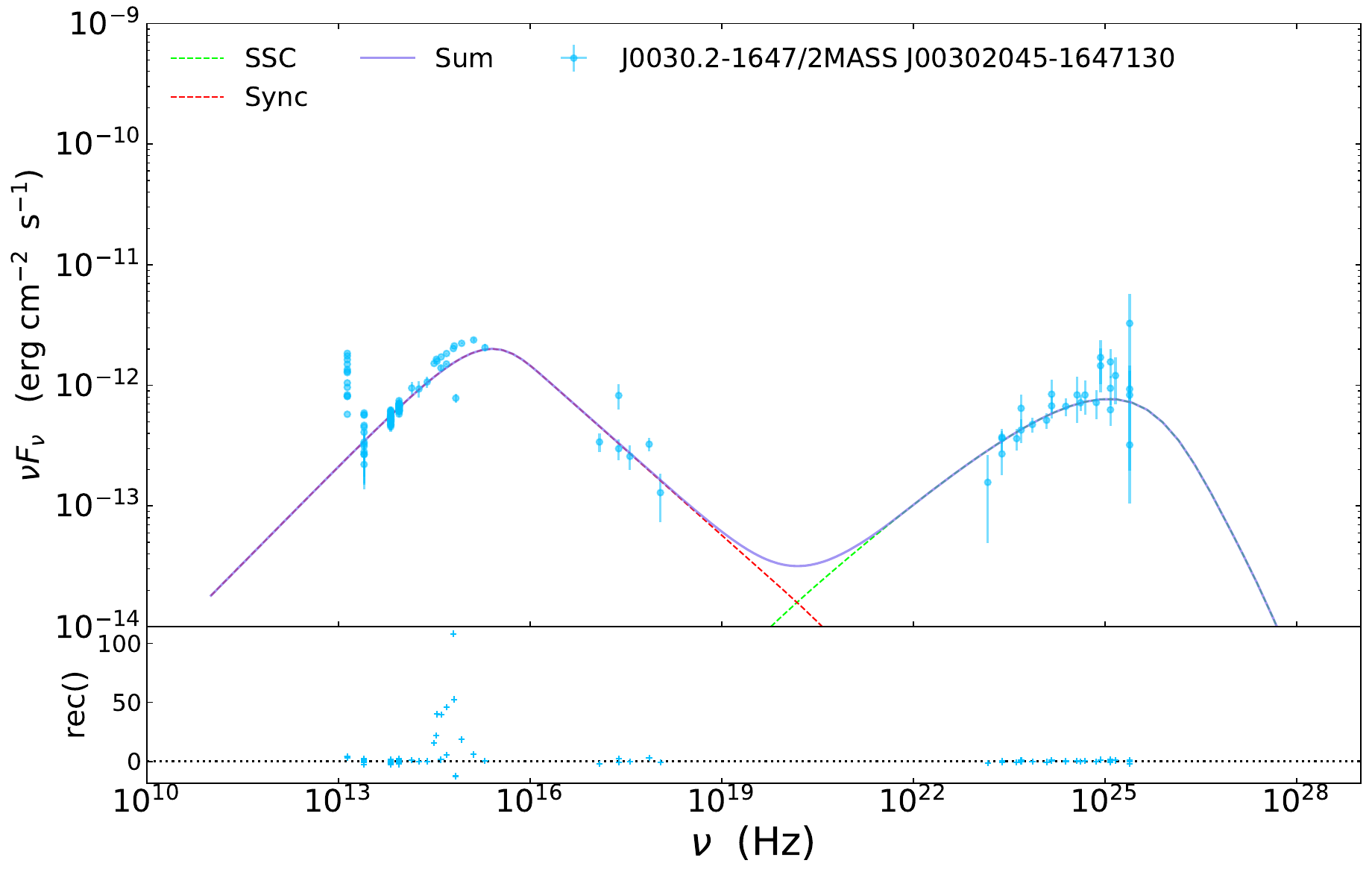}}
		\subfigure{
			\includegraphics[width=0.45\linewidth]{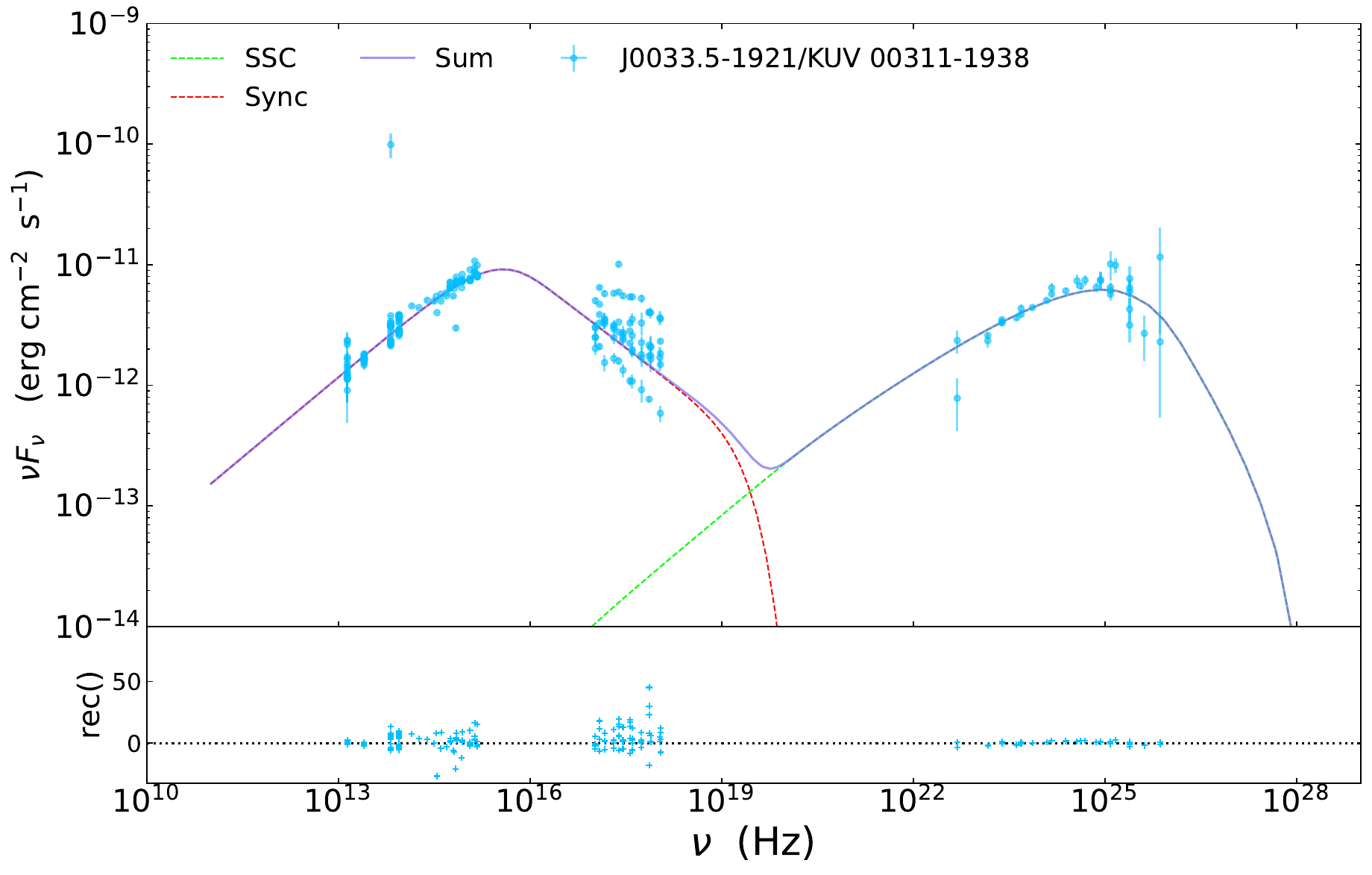}}
	\end{adjustwidth}
	
	\caption{Example of broadband SEDs of J0013.9-1854 (upper left), J0014.1-5022 (upper right), J0030.2-1647(lower left) and J0033.5-1921(lower right) are modeled using a one-zone model.
		Here the solid line is the best fit for the SEDs, the red dashed represents the emission from the synchrotron process, and the green dashed line represents the emission from the SSC process. (The complete figure set (348 images) is available.)}
	
	\label{fig:1}
	
\end{figure*}

\begin{deluxetable*}{ccc}
	\tablenum{3}
	\tablecaption{Fit Range and the Default Values of the
		 Parameters
		 }
	\label{tab:3}
	\tablewidth{0pt}
	\tablehead{
		\colhead{Name} & \colhead{Parameter Type} & \colhead{Fit Range or the Default values}
	}
	\decimalcolnumbers
	\startdata
	$\delta$ & Doppler factor& [5, 50] \\
	B(G) & Magnetic field&[$10^{-4}$, 10] \\
	$\gamma_{\min}$ & Minimum Lorentz factor& [2, $10^4$] \\
    $\gamma_{b}$ & Broken Lorentz factor &[10,$10^7$] \\
    $R_H$(cm) &Position of the region & $10^{17a}$ \\
    $NH\_cold\_to\_rel\_e$&Cold proton to relativistic electron ratio & $0.1^a$ \\
	$\gamma_{max}$ & Maximum Lorentz factor &$[2 \times 10^3, 10^{8}]^b$ \\	
	$p_1$ &Low-energy electron spectral index& $[1,~3.5]^b$ \\
	$delta\_p$ &Customized parameters& $[0,~3.5]^b$ \\
	\textit{gamma\_break\_frac}&Customized parameters& $[0.001,~0.5]^b$ \\
	\enddata
	\begin{tablenotes}   
		\footnotesize               
		\item[1]a.Default value for \textit{JetSet}.
		\item[2]b.Refine the fit range.    			
	\end{tablenotes}      
\end{deluxetable*}

\subsection{Modeling the  SEDs}
To pursue consistency between objects in statistical analysis, we only consider a simple one-zone lepton Syn+SSC model to fit the observed broadband SEDs.
In SED modeling, the radiation region is taken to be a uniform sphere of radius R. The size of the emission region can be derived from the relation (\citealt{2014Natur.515..376G}):
\begin{equation}R=ct_{var}\delta/(1+z)\end{equation}
where $t_{var}$ is the variability timescale. In this paper $t_{var}=1$ day (\citealt{2013RAA....13..259F,2013MNRAS.430.1324N}), $\delta=(\Gamma(1-\beta cos\theta))^{-1}$ is the Doppler factor, and $\theta$ is the angle between the jet axis and line of sight of the observer.
We denote the Lorentz factor as $\Gamma$.
For blazars, $\sin(\theta)\approx1/\Gamma$ and thus, $\Gamma\simeq\delta$ (\citealt{2014Natur.515..376G}).
The jet's relativistic speed is ${\beta}=\sqrt{1-1/\Gamma^{2}}$.
We assume that the electron distribution $N(\gamma)$ in the jet follows a broken power law distribution (\citealt{2010MNRAS.401.1570T,2012ApJ...752..157Z}) governed by the electron spectral index $p_1$ and $p_2$ as follows:
\begin{equation}N(\gamma)=N_0\begin{cases}\gamma^{-p_1}&\gamma_\mathrm{min}\leqslant\gamma\leqslant\gamma_\mathrm{b},\\\gamma_\mathrm{b}^{p_2-p_1}\gamma^{-p_2}&\gamma_\mathrm{b}<\gamma<\gamma_\mathrm{max},\end{cases}\end{equation}
where $\gamma_{min}$, $\gamma_{b}$, and $\gamma_{max}$ are the minimum Lorentz factor, broken Lorentz factor, and maximum Lorentz factor respectively, $N_0$ is the normalization constant in units of 1 $\mathrm{cm}^{-3}$.

Our goal is mainly to reproduce the "average" SED of the source, and it is worth noting that some sources, especially in the X-ray band, show a very large dispersion (due to flare activity), whereas other energy bands collect more dispersed samples. It is possible to obtain an average SED, which is affected by flares in some bands, and sporadic sampling in other bands.

In the \textit{JetSet}, we fit SEDs using the \textit{iminuit} module and determine the best-fitting parameter based on the size of $\chi^{2}/dof$. First, we import the multiband data we collected and add systematics (the size is 0.1 and the range is [$10^6 \sim 10^{29}$]). As we all know, due to the overlapping of different instruments and to snapshots at different times, some points have multiple values, however, this is not a problem for \textit{JetSet}. So we do not bin/average our multiband data in this work. It is worth noting that if we want to bin/average the multiband data, we can use the \textit{$group\_data$} code on the imported data.  
In the second step we obtain the phenomenological model constraints, we use the \textit{SEDshape} module to perform a binned combination of our collected data, with the ranges for radio to TeV bands as shown in Table \ref{tab:2}. And then we use the \textit{ObsConstrain} module; in this step we do not perform a fit, but we obtain the phenomenological model.  In this step we can select our electron distribution: a variety of electron distributions are available in \textit{JetSet}, including log-parabola (lp), power law (pl), log-parabola with low-energy power law branch (lppl), log-parabola defined by peak energy (lpep), power law with cutoff (plc), broken power law (bkn), power law with superexp cut-off (superexp). For HBLs, lp, pl, lppl, plc and bkn etc can be fitted very well (\citealt{2011ApJ...739...66T,2018ApJ...859...49P,2021ApJ...915...59Z,2022MNRAS.513.1662M}), but in this paper we
choose bkn, which is used in \cite{2010MNRAS.401.1570T}, \cite{2012ApJ...752..157Z}, \cite{2016MNRAS.456.2374T} and \cite{2018MNRAS.478.3855Z}, all of which have achieved good fitting results, and we would like to compare our results with those. 
Third, we fit the phenomenological model with multiband data using the fitting method \textit{iminuit}, in \textit{JetSet}, where each free parameter has a specific physical boundary.
We can use \textit{freeze} to fix specific parameters and \textit{$fit\_range$} to set the fitting range of the remaining parameters to speed up convergence of the fit. Considering all our samples are HBLs and adopting the one-zone lepton model Syn+SSC, we fix the redshift and the distance of the radiation region from the central black hole $R_H$.
We have defined fitting ranges for Doppler factor, magnetic field strength, $\gamma_{min}$, and $\gamma_{b}$ to avoid biased results, ensure output parameters fall within physically acceptable ranges, and enhance the convergence rate.
We have studied various articles (\citealt{2004A&A...413..489M,2004A&A...422..103M,2006A&A...448..861M,2010MNRAS.401.1570T,2010MNRAS.402..497G,2012ApJ...752..157Z,2018MNRAS.478.3855Z,2021ApJ...915...59Z,2023ApJS..268...23F}) and decided to choose as an upper limit the maximum value of those found in the literature; the lower bound of the fitting parameter is obtained using the identical methodology.
For parameters without obvious constraints, we use \textit{JetSet}'s default parameter range. Note that the fit will fail if the default parameter range of \textit{JetSet} is exceeded. The relevant information is shown in Table~\ref{tab:3}.
Finally, we compile all obtained results.
In our study, the ratio of cold protons to relativistic electrons is taken as 0.1  (\citealt{2012MNRAS.424L..26G}).

The fact that we did not set specific constraints on certain parameters led to some extreme cases, thus making the results look less reasonable and physically meaningless. For example, $p_1$ is greater than $p_2$, $\gamma_{max}$ is less than $\gamma_{\mathrm{b}}$, or some values reach extreme value. This is not a specific \textit{JetSet} issue, it is a typical problem of parameters boundaries during the minimization process. The functional dependency helps in preventing this problem. We needed to check our results and we found 56 HBLs with extreme cases. In order to ensure that the parameters obtained from our fit are reasonable and physically meaningful, we need to refit and add functional dependences in the third step of the fitting process. We added \textit{`delta\_p'} using \textit{add\_user\_par} code, with an initial value of 1, we set the $delta\_p$ fit range from 0 to 3.5 and define $p_2 = p_1 + delta\_p$, the fitting range of $p_1$ is [1, 3.5]. Secondly, we adopted the same method to add \textit{`gamma\_break\_frac'}, the initial value is 0.01, the fitting range is [0.001,0.5] and we defined $\gamma_{\mathrm{b}}= \gamma_{max} \times gamma\_break\_frac$, where the fitting range of $\gamma_{max}$ is [$2 \times 10^3, 10^8$].

The peak frequency of the SED's two humps can be obtained in \textit{JetSet} using the \textit{$Jet.get\_spectral\_component\_by\_name()$} method, where \textit{$restframe$ = $'obs'$}; we can obtain the peak frequency in the observer"s coordinate system. Utilizing \textit{$restframe$ = $'src'$}, we can obtain the peak frequency in the comoving coordinate system.
On the other hand, using \textit{$energy\_report()$}, we can obtain the energy density and jet power of every component of each source. In this module, we can get the energy report of the jet model.
This report gives energy densities ($U\_$) (in the rest frames of both the blob end disk), the power of the emitted components in the restframe of the blob ($L\_$), and the power carried by the jet ($jet\_L$) for the radiative components, the electrons, the magnetic fields, and for the cold protons in the jet.
Figure~\ref{fig:1} displays our fitted broadband SEDs.

\section{Results and Discussion}  \label{sec:results}
\subsection{The Distribution}

\begin{figure*}[htbp]
	\centering
	\subfigbottomskip=2pt
	\subfigcapskip=-5pt
	\begin{adjustwidth}{-0.0cm}{1cm}
		\subfigure{
			\includegraphics[width=0.34\linewidth]{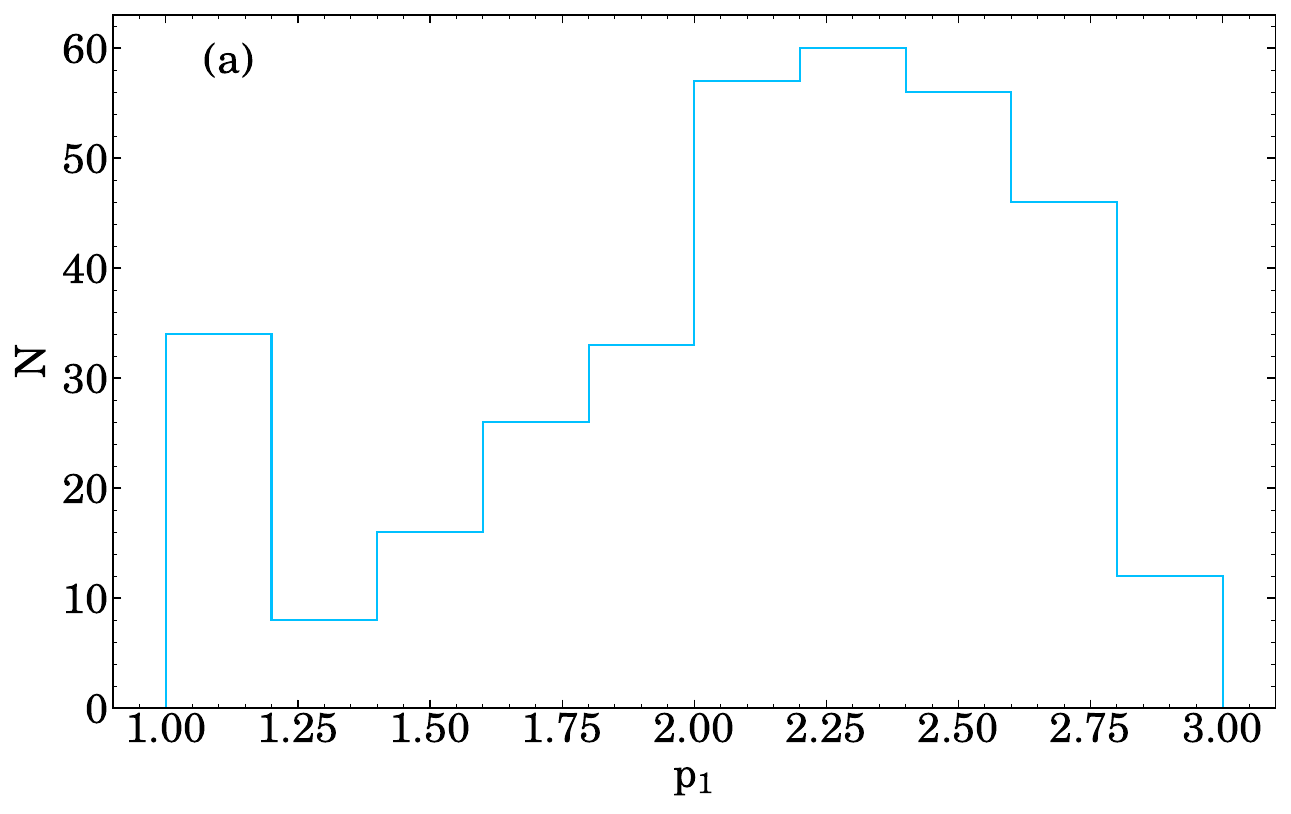}}\hspace{-1mm}
		\subfigure{
			\includegraphics[width=0.34\linewidth]{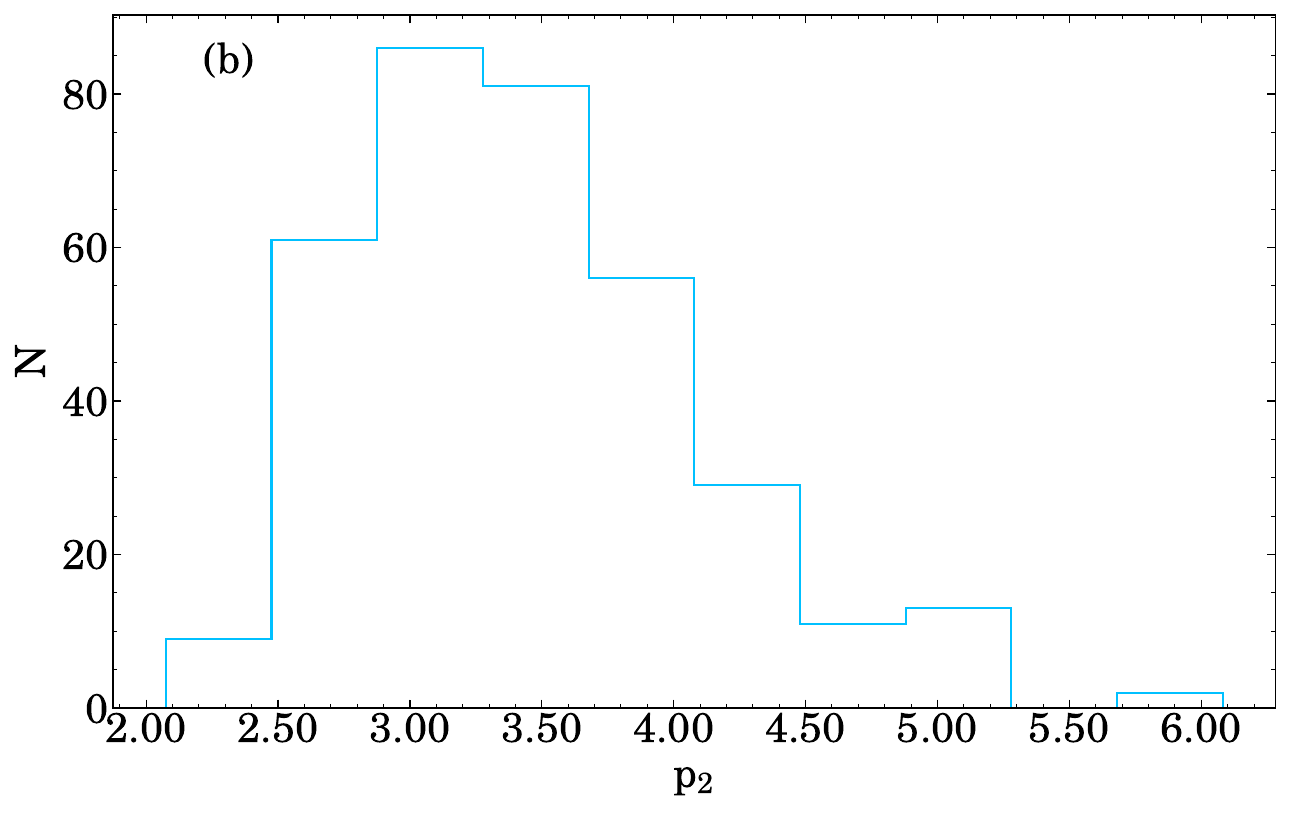}}\hspace{-1mm}
		\subfigure{
			\includegraphics[width=0.34\linewidth]{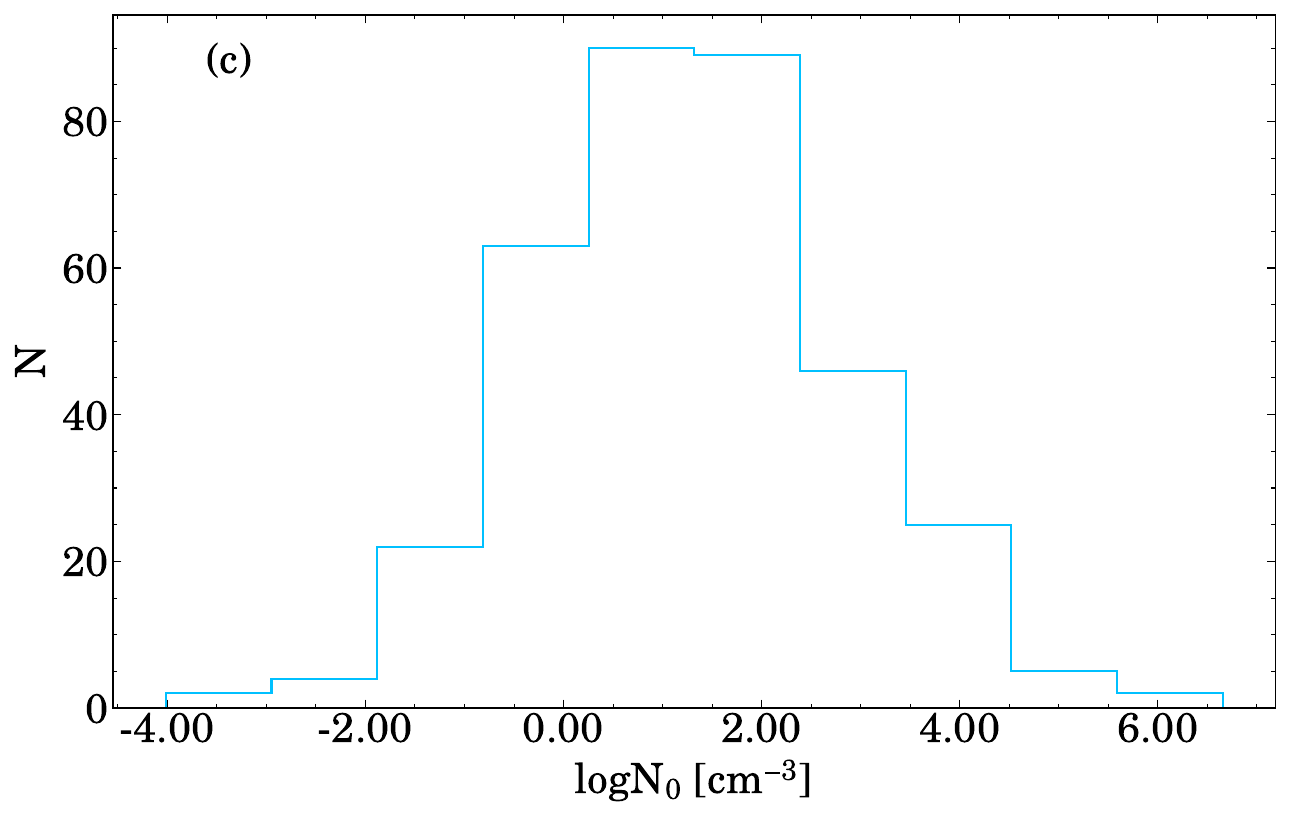}}
		\subfigure{
			\includegraphics[width=0.34\linewidth]{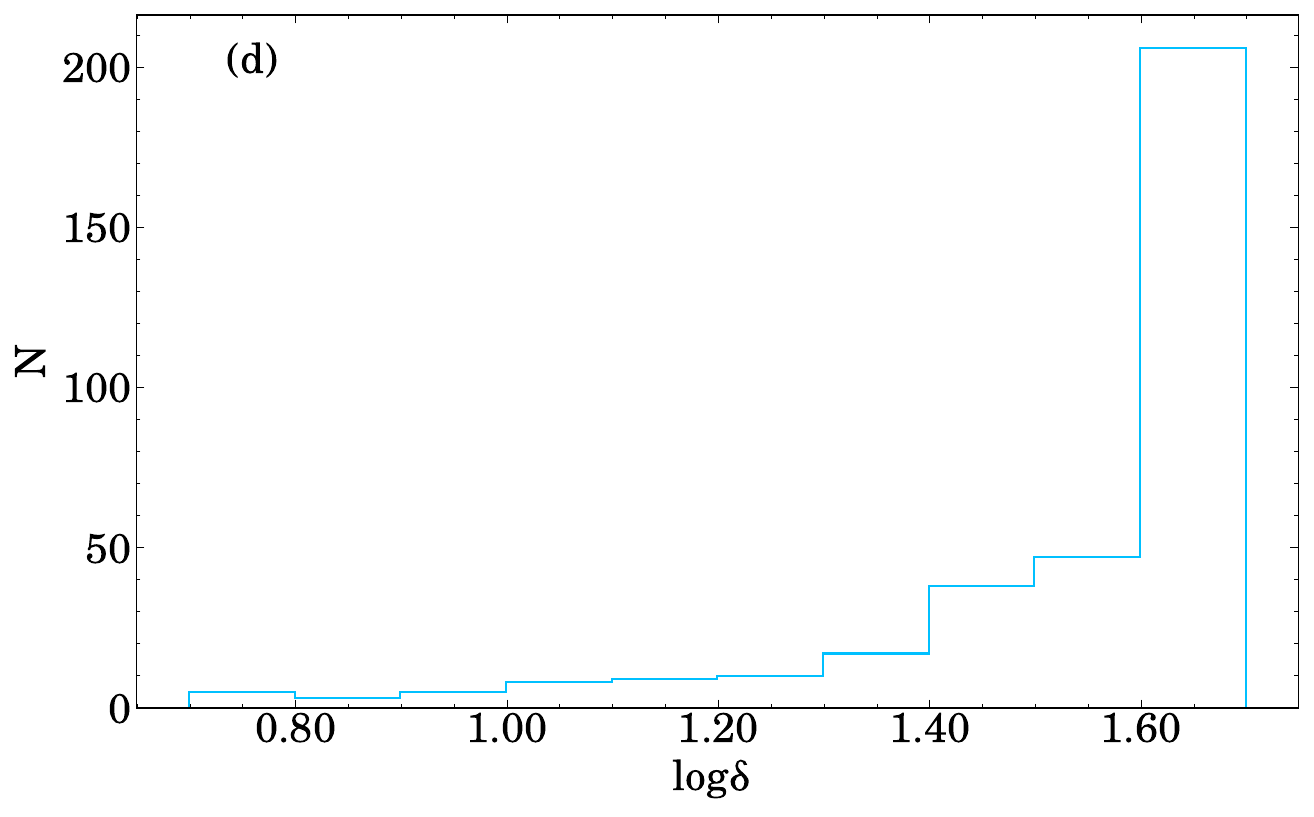}}\hspace{-1mm}
		\subfigure{
			\includegraphics[width=0.34\linewidth]{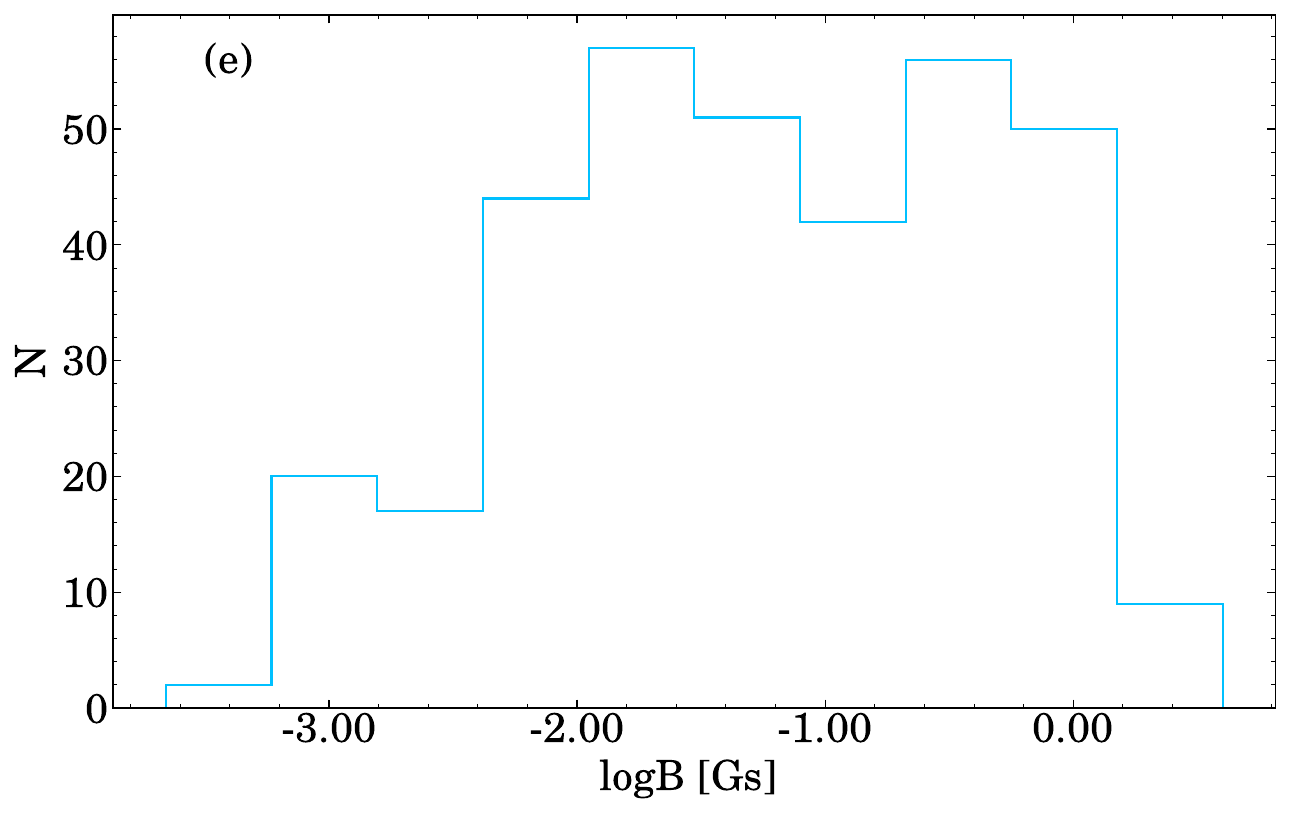}}\hspace{-1mm}
		\subfigure{
			\includegraphics[width=0.34\linewidth]{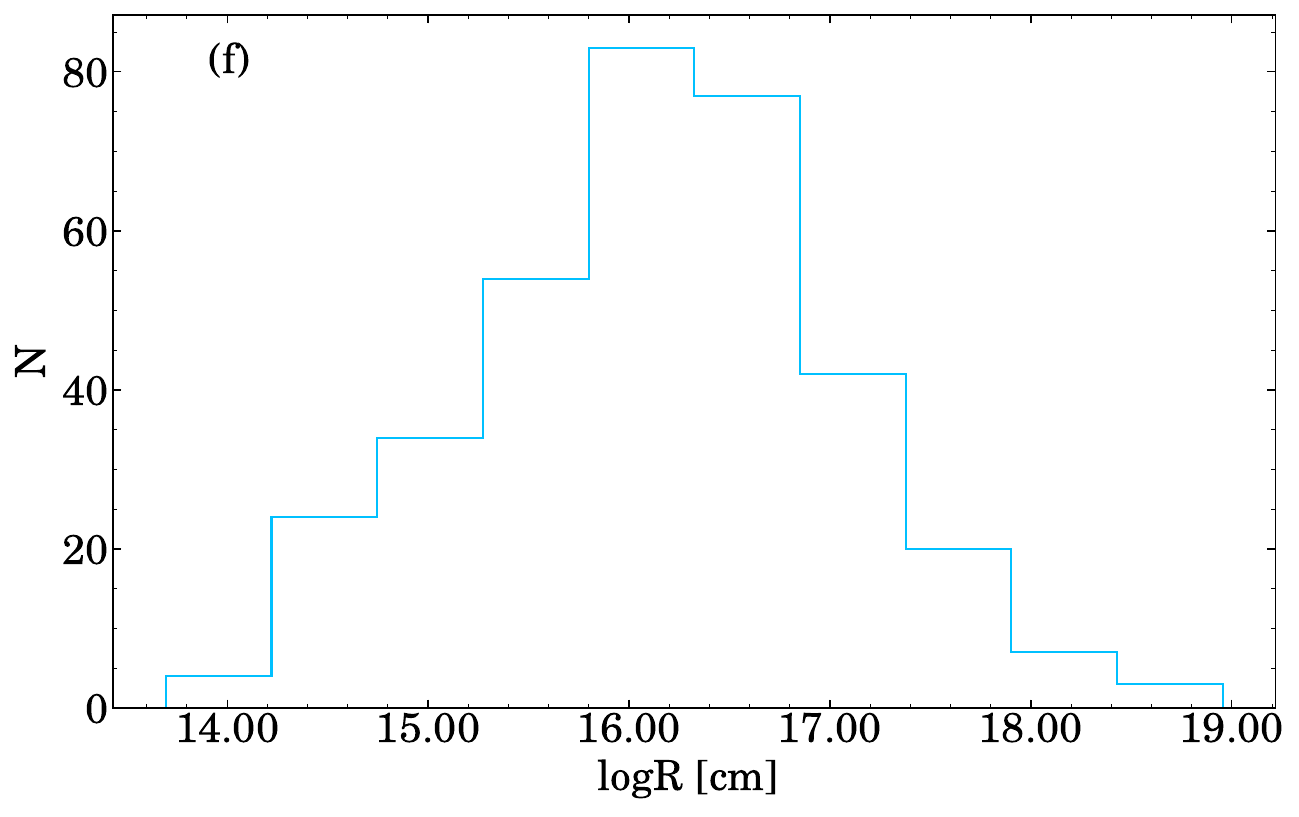}}
		\subfigure{
			\includegraphics[width=0.34\linewidth]{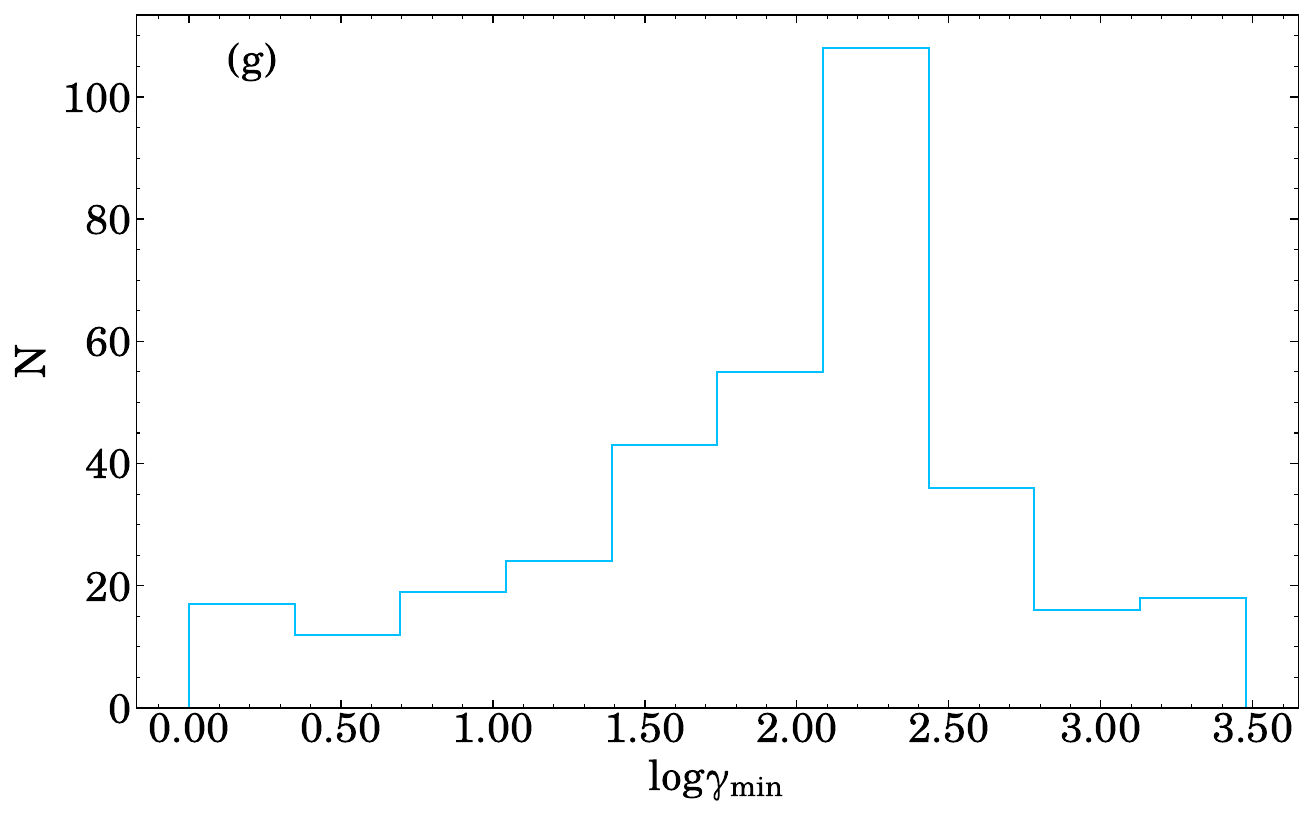}}\hspace{-1mm}
		\subfigure{
			\includegraphics[width=0.34\linewidth]{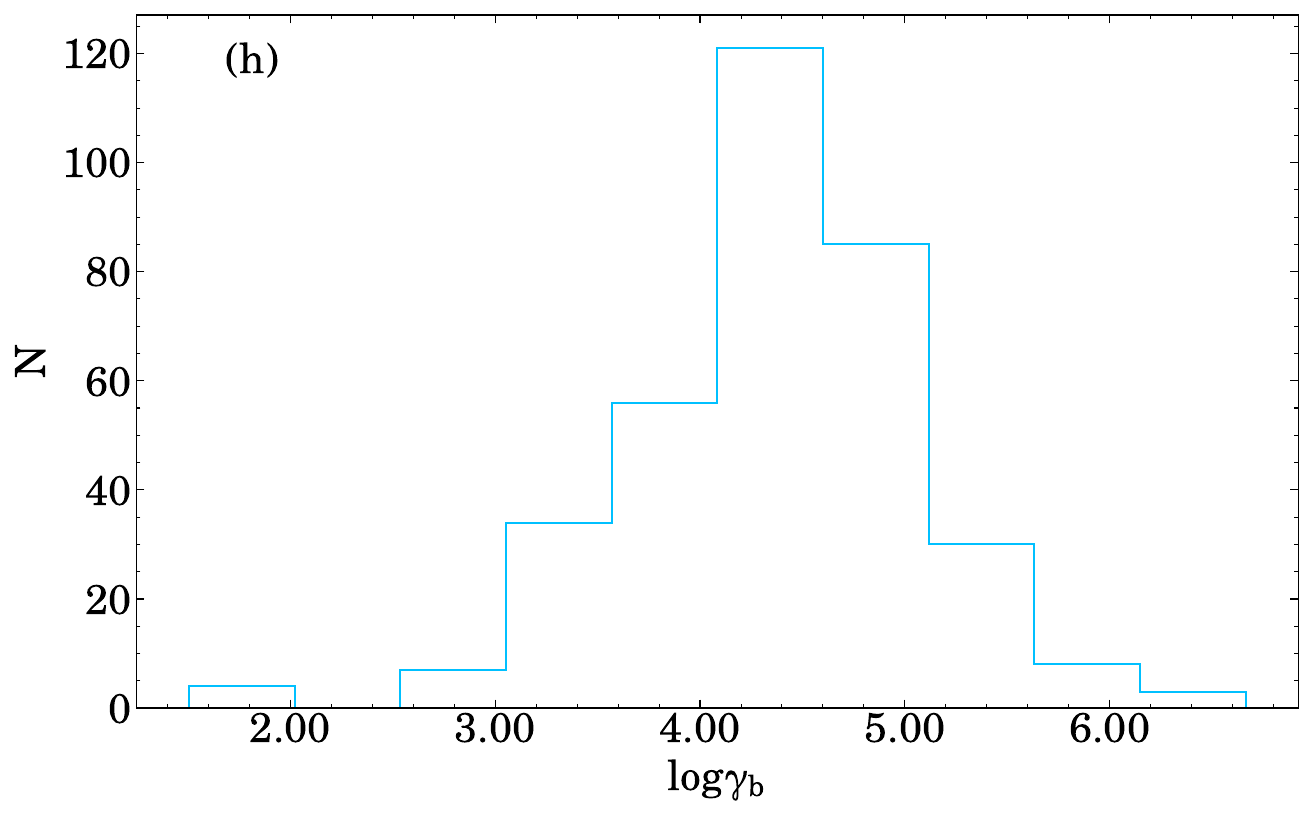}}\hspace{-1mm}
		\subfigure{
			\includegraphics[width=0.34\linewidth]{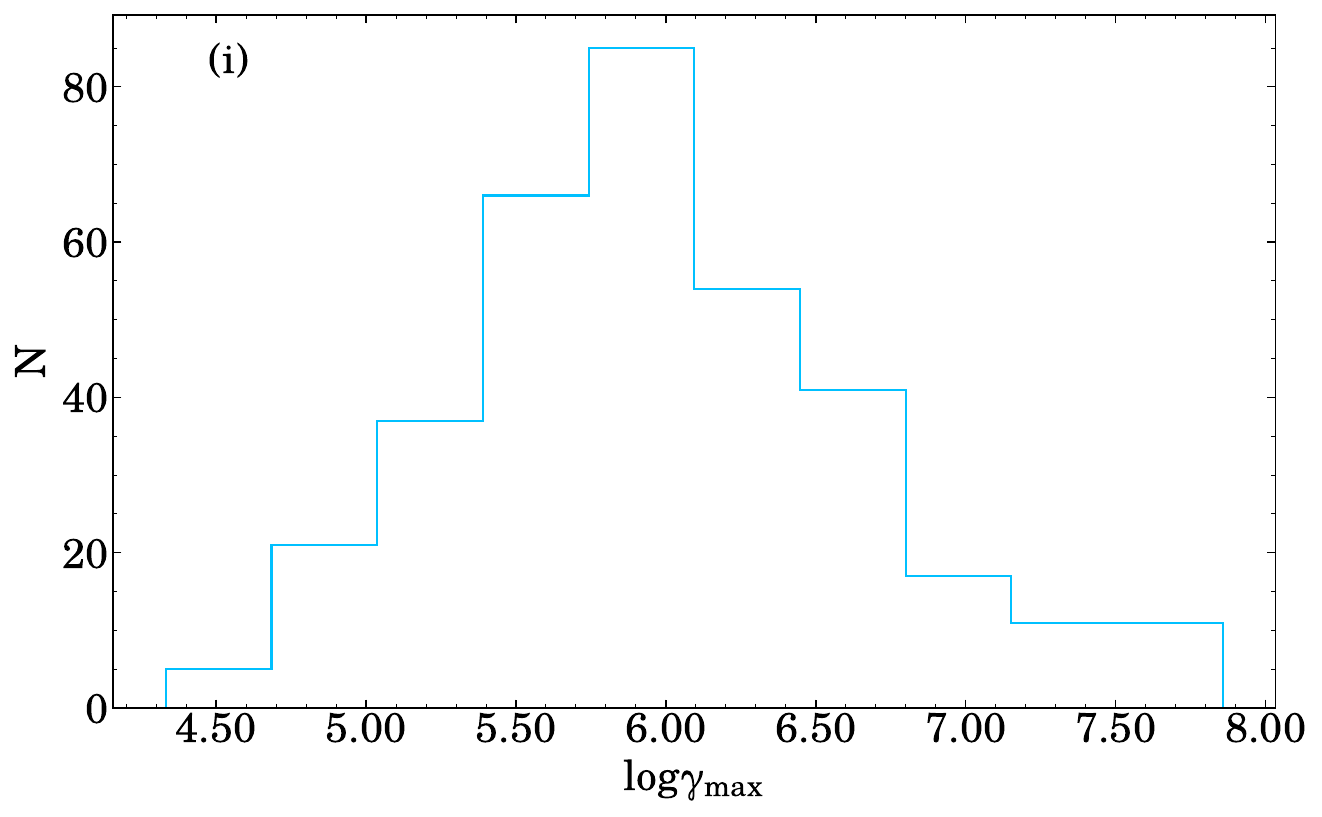}}
	\end{adjustwidth}
	\caption{Distribution of (a, b) the  electron spectral index, (c) the normalization constant, (d) the Doppler factors $\delta$, (e) the magnetic field B, (f) the radiation region R, (g) the minimum Lorentz factor $\gamma_{min}$, (h) the broken Lorentz factor$\gamma_{break}$, and (i) the maximum Lorentz factor $\gamma_{max}$.}
	\label{fig:2}
\end{figure*}

\begin{figure*}[htbp]
	\centering
	\subfigbottomskip=2pt
	\subfigcapskip=-5pt
	\begin{adjustwidth}{-0.0cm}{1cm}
		\subfigure{
			\includegraphics[width=0.34\linewidth]{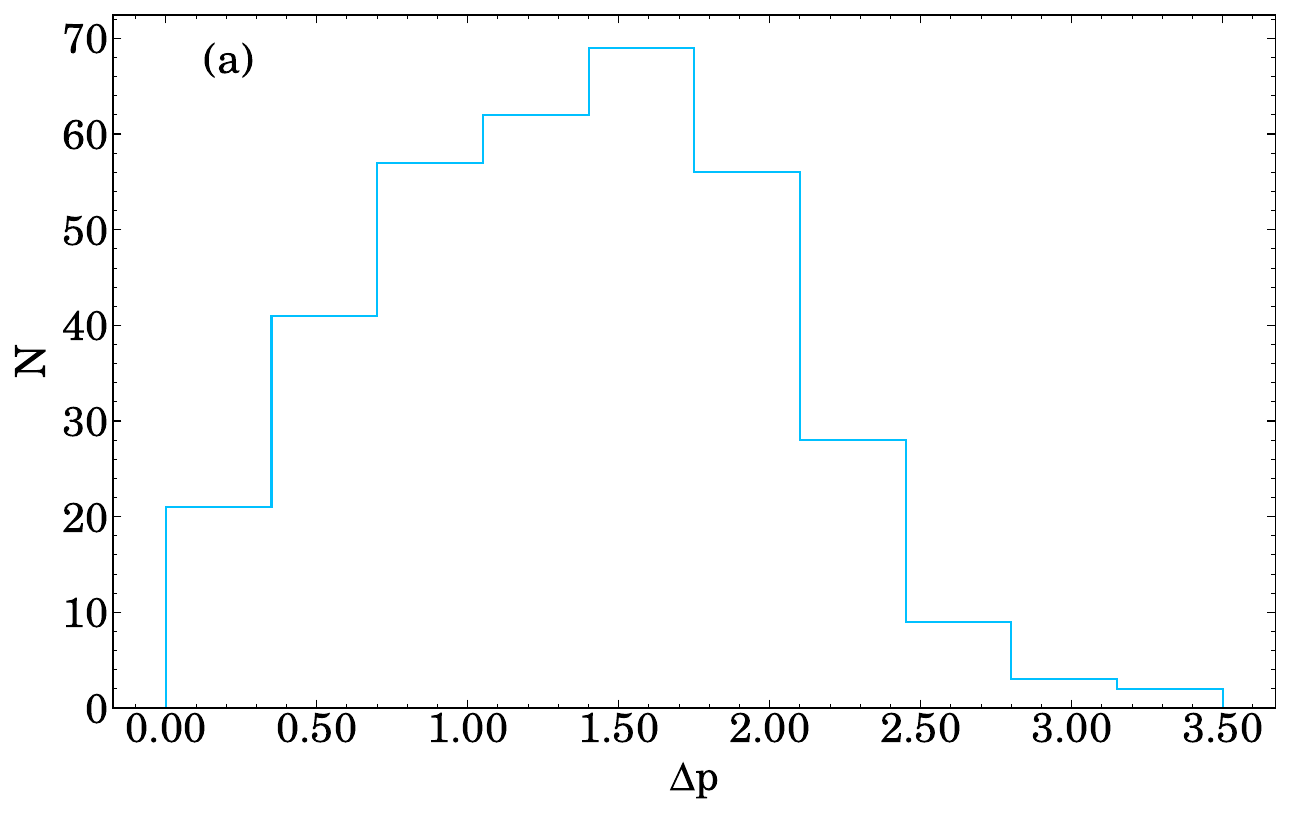}}\hspace{-1mm}
		\subfigure{
			\includegraphics[width=0.34\linewidth]{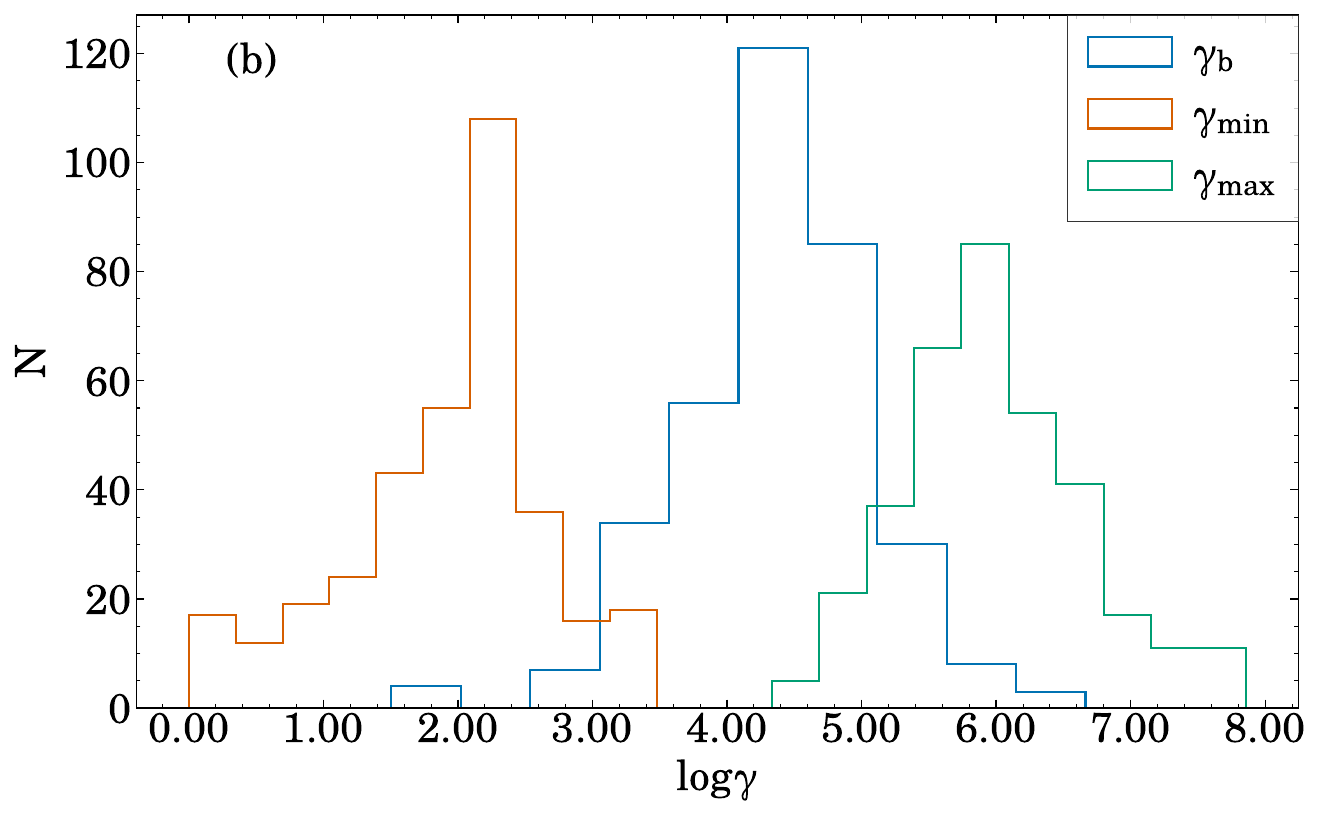}}\hspace{-1mm}
		\subfigure{
			\includegraphics[width=0.34\linewidth]{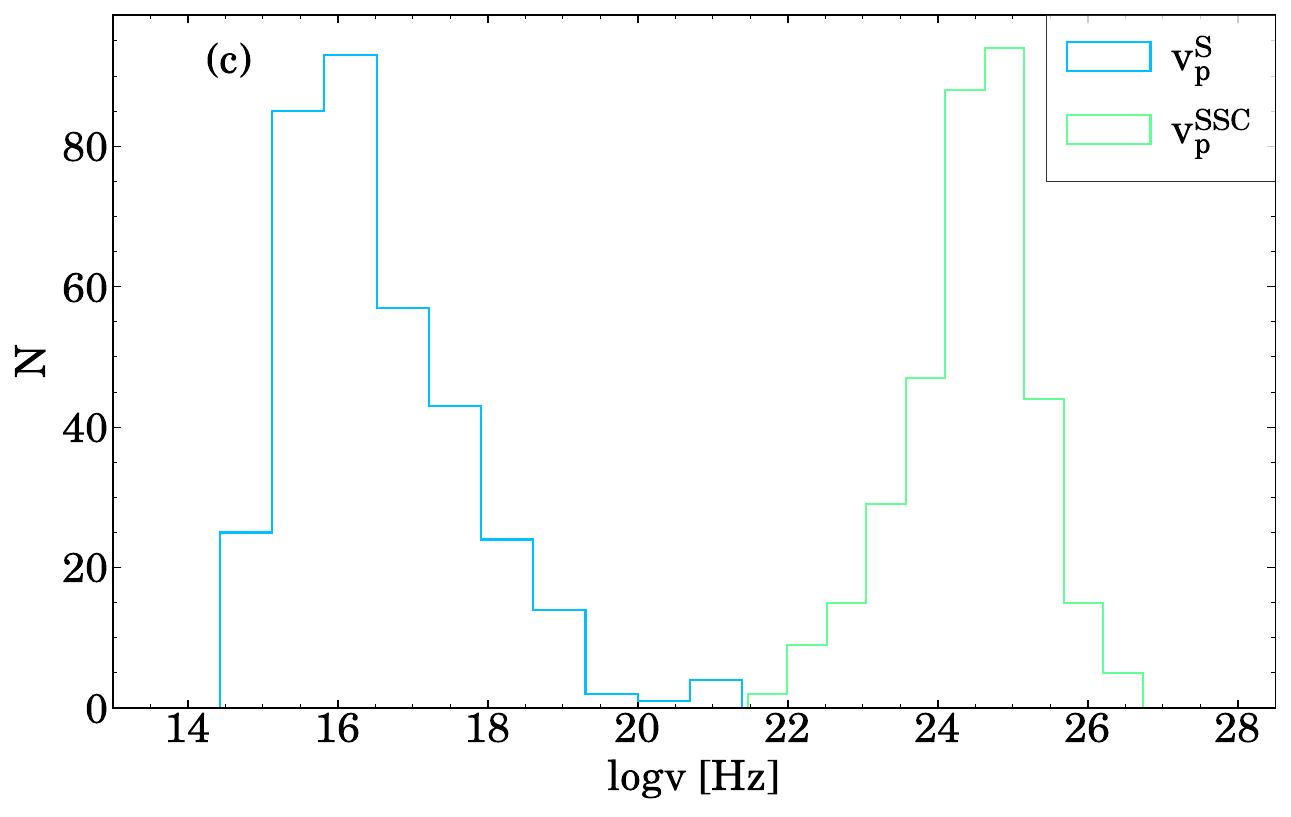}}

	\end{adjustwidth}
	\caption{(a) Distribution of the  $\Delta p$ ($\Delta p = p_2-p_1$). (b) Distributions of $\gamma_{min} (yellow), \gamma_{\mathrm{b}} (blue), \gamma_{max}$ (green). (c) Distribution of the peak frequency, the blue shows the synchrotron radiation peak frequency ($\log v_{\mathrm{p}}^{\mathrm{S}}$), green shows the synchrotron self-Compton peak frequency ($\log v_{p}^{SSC}$).}
	\label{fig:3}
\end{figure*}

\begin{figure*}[!htbp]
	\centering
	\subfigbottomskip=2pt
	\subfigcapskip=-5pt
	\begin{adjustwidth}{-0.3cm}{-1.cm}
		\subfigure{
			\includegraphics[width=0.46\linewidth]{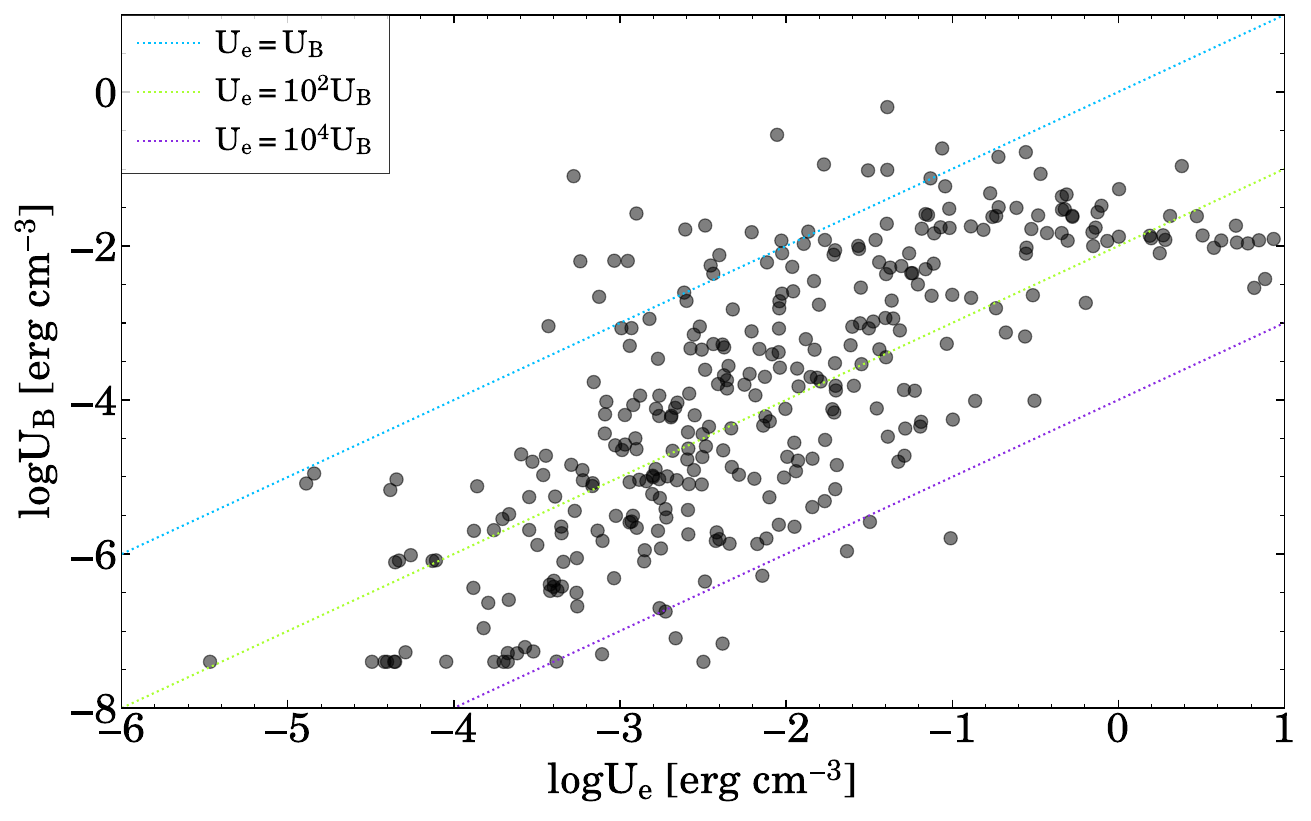}}
		\subfigure{
			\includegraphics[width=0.46\linewidth]{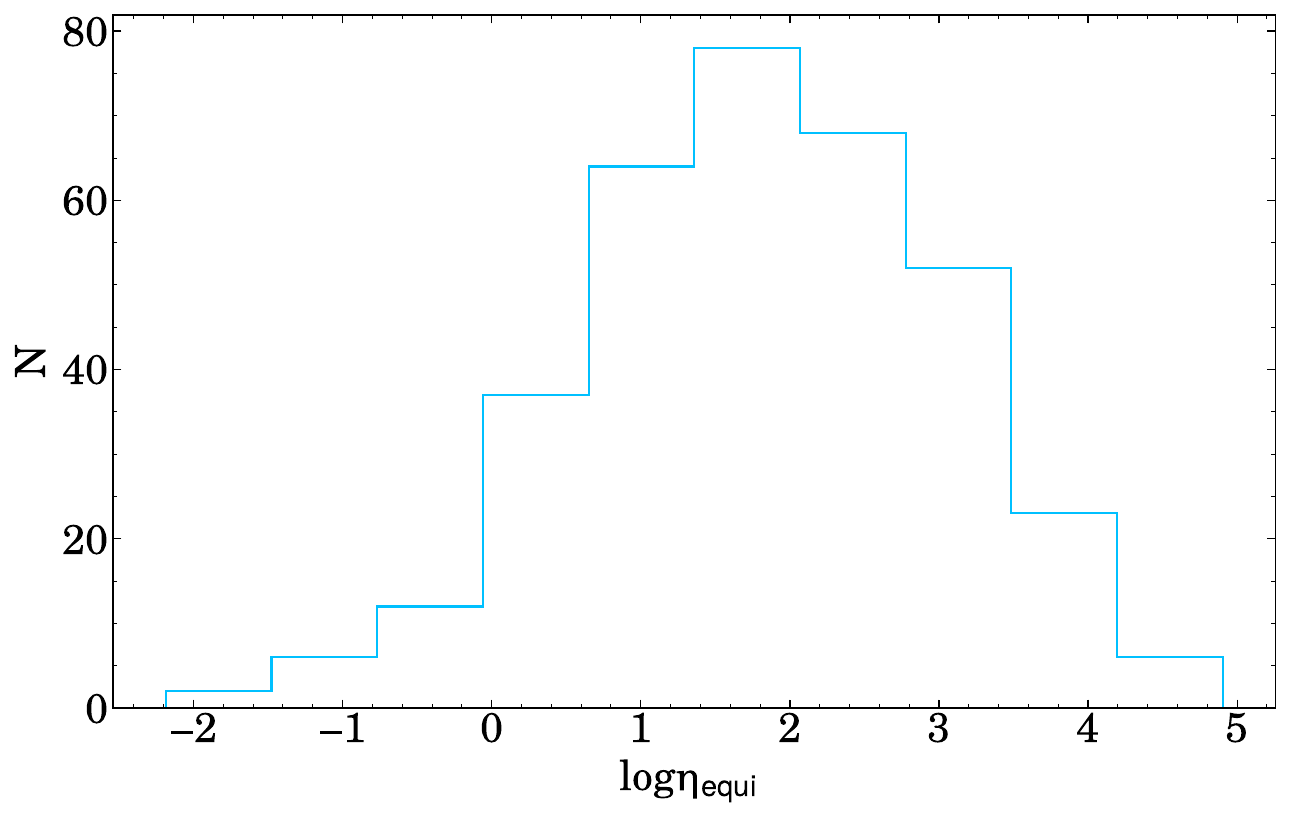}}
	\end{adjustwidth}
\caption{Left panel shows the relation between the magnetic energy density (y-axis) and the electron energy density (x-axis).
The green dotted line represents $U_e=U_B$, the blue dotted line represents  $U_e=10^2U_B$, and the purple dotted line represents  $U_e=10^4U_B$.
Right panel shows the distribution of the equipartition coefficient $\eta_{\mathrm{equi}}$.}
	\label{fig:4}
\end{figure*}

We present the best-fitting values of the parameters in Table~\ref{tab:1}. In Figure~\ref{fig:2} (a-i), we display the parameter distribution of jet of the HBLs.
The range of the Doppler factor of HBLs is $\begin{aligned}0.70\leq\log\delta\leq1.70\end{aligned}$, with a mean value $\langle\log\delta_{}\rangle\simeq1.55$.
The magnetic field is in the range $\begin{aligned}-3.66 Gs\leq\log B\leq0.60 Gs\end{aligned}$, with a mean value of $\langle\log B\rangle\simeq-1.25Gs$. The size of emission region is in the range  $\begin{aligned}13.69\operatorname{cm}\leq\log R\leq18.95\operatorname{cm}\end{aligned}$, with a mean value of $\langle\log R\rangle\simeq16.14\operatorname{cm}$.
The range of the low-energy electron spectral index parameter $p_1$ is $1\begin{aligned}\leq p_1\leq3\end{aligned}$with a mean value is $\langle p_1\rangle\simeq2.11$.
The range of the high-energy electron spectral index parameter $p_2$ range is $2.07\begin{aligned}\leq p_2\leq6.08\end{aligned}$ with a mean value of $\langle p_2\rangle\simeq3.44$.
In Figure \ref{fig:3} (a), we plot a histogram of $\Delta p$ ($\Delta p = p_2-p_1$), the mean value of  $\langle \Delta p\rangle$ in our sample is 1.34 and the median is 1.36. Since in the slow cooling case. We can deduce $\Delta p = 1$, but  $\Delta p = s-1$ in the case of fast cooling, we can deduce, with the spectrum index of injected particles s=2.5 \citep{2018MNRAS.478.3855Z}. Therefore, it can be seen that the result in this
context is closer to the fast cooling scenario.
The range of $\gamma_{b}$ is $\begin{aligned}1.50\leq\log\gamma_{b}\leq6.66\end{aligned}$ with a mean value of $\langle\log\gamma_{b}\rangle\simeq4.38$.
In addition, in Figure \ref{fig:3} (b), we plot histograms of $\gamma_{\mathrm{b}}, \gamma_{min}$, and $\gamma_{max}$, from which we can see that the distributions of the three are similar; the mean value of  $\langle gamma\_break\_frac \rangle$ in our sample is 0.08 and the median is 0.02. There are some larger values of $\gamma_{max} $ in our sample. Our results are similar to those of \cite{2021Ap&SS.366...36D} and \cite{2023Galax..12....2D}, who both utilized the same electron distribution model and larger $\gamma_{max}$ values in their SED fitting. We infer that this portion of these sources may be candidates for extreme high-energy peaked BL Lac objects, but the specific details need to be confirmed by obtaining more details, and we leave this to subsequent work.

Compared to findings from other authors (\citealt{2010MNRAS.402..497G,2012ApJ...752..157Z}), the parameter range in our  results aligns closely, but disparities exist in the mean values.
These differences could stem from several factors.
(1) Our study focuses explicitly on the HBL subclass of BL Lac objects, whereas the dataets of other authors predominantly comprise HBLs but also include various other BL Lac objects.
(2) The multiband data employed in our analysis represent an average state. (3) Variations in the methodologies used for calculating parameter may also contribute to differences in the rresult.

The main acceleration processes of blazar jets are assumed to be shock acceleration and stochastic acceleration. Since the acceleration rate of these two processes depends on the nature of the diffusion of particles into the jet medium, a simple judgment can be made by means of a power-law index \citep{2022MNRAS.514.3074B}. In the case of strong electrostatic/shock acceleration, we expect to find that the high-energy power-law index is in the range of [2, 3] (\citealt{2016ApJ...833..157K,2019ApJ...873....7Z}). The case of strong electrostatic/shock acceleration accounts for $33 \%$ of HBLs in our samples, while other acceleration mechanisms may be responsible for the remainder. However, a more detailed study/simulation of the process of HBL jet acceleration is needed in future work if a more accurate understanding of the HBL acceleration mechanism is to be obtained.

Figure \ref{fig:3} (c) shows the distribution of the synchrotron peaked frequency and the synchrotron self-Compton peaked frequency ($v_{\mathrm{p}}^{\mathrm{SSC}}$) of HBLs.
We find that the range of $\log v_{\mathrm{p}}^{\mathrm{S}}$ is $15.00 \mathrm{Hz}\leq\log v_{\mathrm{p}}^{\mathrm{S}}\leq21.39 \mathrm{Hz}$with a mean value of $\langle\log v_{\mathrm{p}}^{\mathrm{S}}\rangle\simeq16.52 \mathrm{Hz}$, and the range of $\log v_{p}^{SSC}$ is $21.46 \mathrm{Hz}\leq\log v_{\mathrm{p}}^{\mathrm{SSC}}\leq26.73 \mathrm{Hz}$, with a mean value of $\langle\log v_{\mathrm{p}}^{\mathrm{SSC}}\rangle\simeq24.44 \mathrm{Hz}$.

\begin{deluxetable*}{cccccc}
	\tablenum{4}
	\tablecaption{Mass of the Central Black Hole and the Accretion Disk Luminosity
		}
\label{tab:4}
	\tablewidth{0pt}
	\tablehead{
		\colhead{$\rm 4FGL\;Name$} & \colhead{log $M_{BH}$} & \colhead{log $L_{disk}$}&\colhead{$\rm 4FGL\;Name$} & \colhead{log $M_{BH}$} & \colhead{log $L_{disk}$}
	}
	\decimalcolnumbers
	\startdata
	{$ \rm J0013.9-1854$} & 9.65  & 43.27 & {$ \rm J1136.4+6736$} & 8.67  & 43.49 \\
	{$ \rm J0022.0+0006$} & 8.02  & 43.79 & {$ \rm J1140.5+1528$} & 8.70  & 43.97 \\
	{$ \rm J0059.3-0152$} & 8.63  & 43.52 & {$ \rm J1145.5-0340$} & 8.39  & 43.28 \\
	{$ \rm J0115.8+2519$} & 9.03  & 44.53 & {$ \rm J1149.4+2441$} & 8.21  & 44.32 \\
	{$ \rm J0121.8-3916$} & 8.58  & 44.39 & {$ \rm J1154.0-0010$} & 8.13  & 43.75 \\
	{$ \rm J0152.6+0147$} & 9.34  & 42.92 & {$ \rm J1212.0+2242$} & 8.16  & 44.50 \\
	{$ \rm J0201.1+0036$} & 8.19  & 43.93 & {$ \rm J1221.3+3010$} & 9.00   & 44.17 \\
	{$ \rm J0232.8+2018$} & 10.08 & 43.18 & {$ \rm J1224.4+2436$} & 8.29  & 44.11 \\
	{$ \rm J0237.6-3602$} & 8.31  & 44.66 & {$ \rm J1251.2+1039$} & 7.62  & 43.89 \\
	{$ \rm J0238.7+2555$} & 9.63  & 44.56 & {$ \rm J1253.8+0327$} & 8.49  & 42.65 \\
	{$ \rm J0250.6+1712$} & 9.47  & 44.23 & {$ \rm J1256.2-1146$} & 8.94  & 43.06 \\
	{$ \rm J0304.5-0054$} & 9.16  & 44.68 & {$ \rm J1257.6+2413$} & 8.56  & 43.33 \\
	{$ \rm J0305.1-1608$} & 9.24  & 44.10  & {$ \rm J1305.9+3858$} & 8.91  & 44.07 \\
	{$ \rm J0316.2-2608$} & 9.42  & 44.66 & {$ \rm J1326.1+1232$} & 8.75  & 43.70 \\
	{$ \rm J0325.5-5635$} & 9.10   & 42.84 & {$ \rm J1340.8-0409$} & 9.60  & 43.96 \\
	{$ \rm J0326.2+0225$} & 9.21  & 43.51 & {$ \rm J1341.2+3958$} & 8.91  & 43.45 \\
	{$ \rm J0338.1-2443$} & 9.72  & 43.73 & {$ \rm J1348.9+0756$} & 8.90  & 43.70 \\
	{$ \rm J0339.2-1736$} & 8.98  & 43.39 & {$ \rm J1400.2-4010$} & 8.93  & 43.20 \\
	{$ \rm J0416.9+0105$} & 8.40   & 44.15 & {$ \rm J1406.9+1643$} & 8.90  & 44.77 \\
	{$ \rm J0505.6+0415$} & 9.32  & 44.26 & {$ \rm J1410.3+6058$} & 8.62  & 43.99 \\
	{$ \rm J0558.0-3837$} & 9.81  & 44.52 & {$ \rm J1411.8+5249$} & 8.82  & 42.78 \\
	{$ \rm J0710.4+5908$} & 9.75  & 42.75 & {$ \rm J1416.1-2417$} & 9.21  & 43.48 \\
	{$ \rm J0744.1+7434$} & 9.94  & 44.15 & {$ \rm J1417.9+2543$} & 8.17  & 43.59 \\
	{$ \rm J0809.6+3455$} & 8.80   & 43.00 & {$ \rm J1428.5+4240$} & 8.59  & 43.59 \\
	{$ \rm J0809.8+5218$} & 8.55  & 43.85 & {$ \rm J1438.6+1205$} & 7.91  & 45.05 \\
	{$ \rm J0814.4+2941$} & 8.58  & 44.96 & {$ \rm J1439.3+3932$} & 8.79  & 44.65 \\
	{$ \rm J0830.0+5231$} & 8.79  & 43.53 & {$ \rm J1439.9-3953$} & 8.80  & 43.83 \\
	{$ \rm J0837.3+1458$} & 7.74  & 43.43 & {$ \rm J1442.6-4623$} & 9.25  & 43.25 \\
	{$ \rm J0850.5+3455$} & 8.67  & 43.82 & {$ \rm J1442.7+1200$} & 8.74  & 43.64 \\
	{$ \rm J0912.9-2102$} & 9.53  & 43.93 & {$ \rm J1508.8+2708$} & 8.30  & 43.95 \\
	{$ \rm J0916.7+5238$} & 8.60   & 43.66 & {$ \rm J1518.6+4044$} & 8.25  & 42.65 \\
	{$ \rm J0917.3-0342$} & 9.34  & 44.09 & {$ \rm J1626.3+3514$} & 8.82  & 44.47 \\
	{$ \rm J0930.5+4951$} & 8.87  & 43.64 & {$ \rm J1640.9+1143$} & 9.71  & 43.06 \\
	{$ \rm J0940.4+6148$} & 8.63  & 43.66 & {$ \rm J1653.8+3945$} & 9.91  & 43.30 \\
	{$ \rm J0946.2+0104$} & 7.70   & 43.43 & {$ \rm J1744.0+1935$} & 9.69  & 43.28 \\
	{$ \rm J1010.2-3119$} & 9.97  & 43.57 & {$ \rm J1814.0+3828$} & 9.14  & 43.78 \\
	{$ \rm J1023.8+3002$} & 9.06  & 44.43 & {$ \rm J1954.9-5640$} & 8.37  & 44.08 \\
	{$ \rm J1033.5+4221$} & 8.65  & 43.52 & {$ \rm J2000.0+6508$} & 9.07  & 42.89 \\
	{$ \rm J1046.8-2534$} & 9.94  & 43.77 & {$ \rm J2158.8-3013$} & 8.91  & 43.51 \\
	{$ \rm J1049.7+5011$} & 8.49  & 44.01 & {$ \rm J2159.1-2840$} & 8.31  & 44.02 \\
	{$ \rm J1057.8-2754$} & 9.36  & 43.69 & {$ \rm J2220.5+2813$} & 8.62  & 43.31 \\
	{$ \rm J1104.4+3812$} & 9.05  & 42.18 & {$ \rm J2232.8+1334$} & 8.32  & 43.55 \\
	{$ \rm J1112.4+1751$} & 8.46  & 44.45 & {$ \rm J2250.0+3825$} & 9.44  & 43.31 \\
	{$ \rm J1117.0+2013$} & 8.51  & 43.71 & {$ \rm J2314.0+1445$} & 9.11  & 43.55 \\
	{$ \rm J1117.2+0008$} & 8.79  & 44.35 & {$ \rm J2319.1-4207$} & 9.41  & 43.30 \\
	{$ \rm J1130.5-3137$} & 9.24  & 43.43 & {$ \rm J2322.7+3436$} & 9.55  & 42.80 \\
	{$ \rm J1133.8-2048$} & 8.71  & 42.67 & {$ \rm J2343.6+3438$} & 8.47  & 44.19 \\
	\enddata
	
	\tablecomments{Columns (1) and (4) is the 4FGL name of sources; columns (2) and (5) is the mass of the central black hole, in $M_{\odot}$;columns (3) and (6) is the accretion disk luminosity, in erg $s^{-1}$ }	
\end{deluxetable*}

\subsection{Electron and Magnetic Field Energy Densities} \label{sec:4.2}

\begin{figure*}[!htbp]
	\centering
	\subfigbottomskip=2pt
	\subfigcapskip=-5pt
	\begin{adjustwidth}{-0.3cm}{-1cm}	
	\subfigure{
		\includegraphics[width=0.46\linewidth]{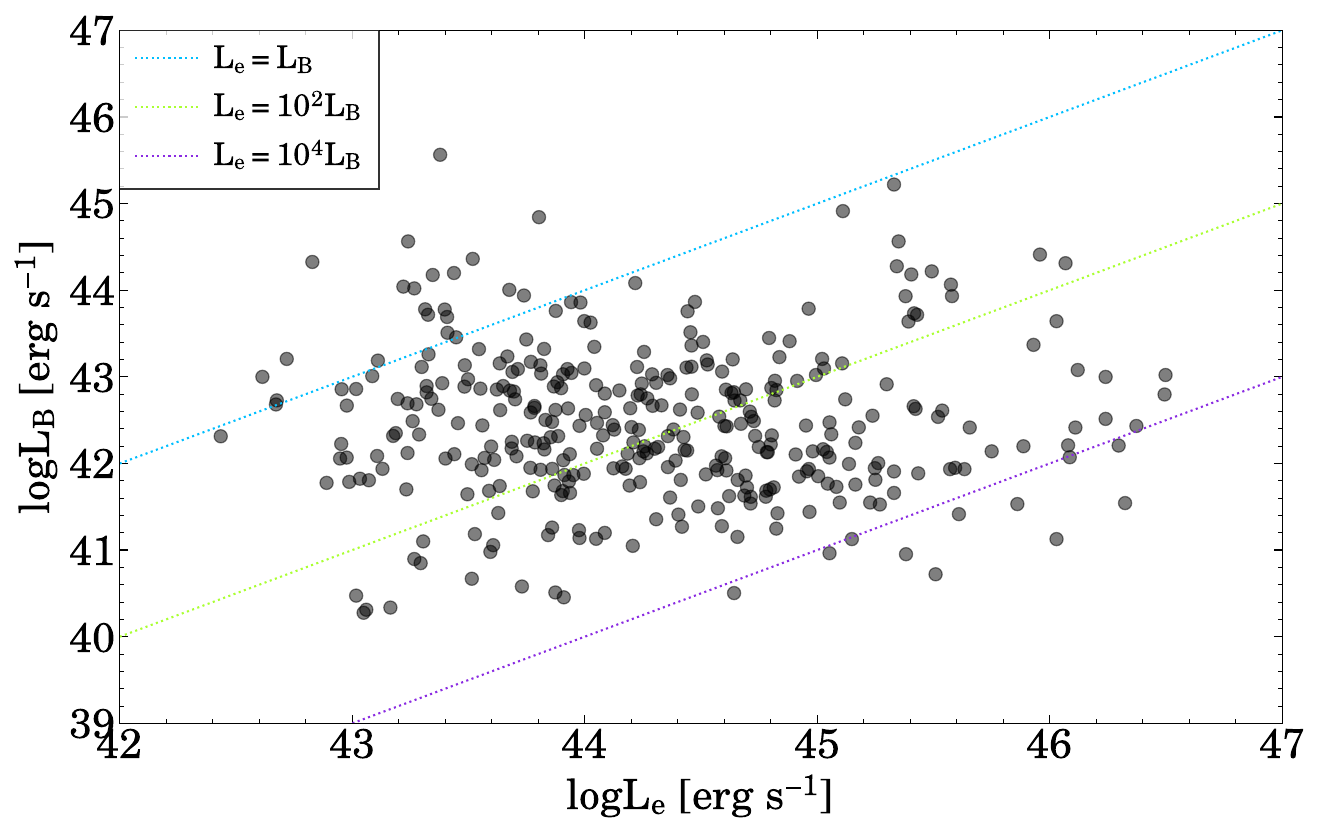}}
	\subfigure{
		\includegraphics[width=0.46\linewidth]{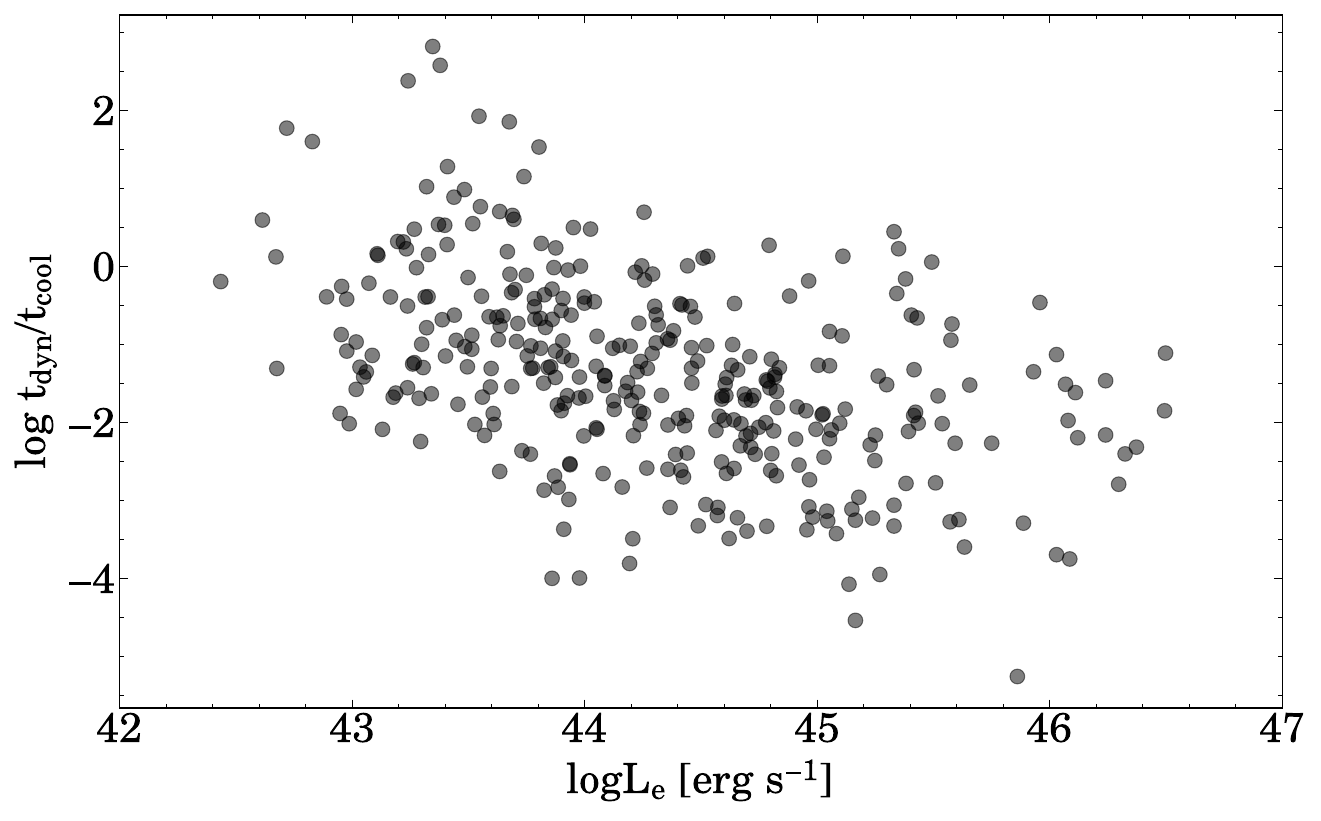}}
	\end{adjustwidth}
\caption{Left panel: the relationship between the magnetic power (y-axis) and the electron power (x-axis).
The green dotted line represents $L_e=L_B$, the blue dotted line represents  $L_e=10^2L_B$, and the purple dotted line represents  $L_e=10^4L_B$.
Right panel: $t_{dyn}/t_{cool}$ as a function of the $L_e$.}
	\label{fig:5}
\end{figure*}
The blazar jet contains relativistic electron energy density and magnetic field energy density.
In the system of a black hole-based jet, the particle-field relations depend on uncertain mechanisms of jet formation, particle acceleration, and radiation, it is natural that systems with interacting components often tend to equipartition (\citealt{2014ApJ...782...82D}).
Nevertheless, \cite{2018ApJ...853....6L} contend that the jet does not consistently maintain equipartition between the particles and magnetic field.
Thus we introduce an equipartition coefficient $\eta_{\mathrm{equi}}$ (\citealt{2017ApJS..228....1Z,2018MNRAS.478.3855Z}), i.e.,
\begin{equation}\eta_{\mathrm{equi}}=\frac{U_{\mathrm{e}}}{U_{B}}\end{equation} where $U_{\mathrm{e}}$ is the electron energy density and $U_{\mathrm{B}}$ is the magnetic field energy density (\citealt{2010MNRAS.402..497G}),
\begin{equation}U_{\mathrm{e}}=m_{\mathrm{e}}c^{2}\int N(\gamma)\gamma d\gamma\end{equation}
\begin{equation}U_{\mathrm{B}}=B^{2}/8\pi\end{equation}

 Figure~\ref{fig:4} indicates the distribution and comparison of the magnetic and electron energy densities on the left, and the distribution of the equipartition coefficient on the right.
 We find that:
 (1) under the simple  one-zone lepton Sync+SSC model, approximately $93\%$ of HBLs in our samples are identified, with $U_{B}<<U_{e}$;
 (2) the range of the equipartition coefficient $\log\eta_{\mathrm{equi}}$ is from -2.19 to 4.90, and its average is $\langle\log\eta_\text{equi}\rangle\simeq1.82$.
 There exists a substantial gap between the electron energy density and the magnetic energy density in our HBL samples, which may be due to the smaller magnetic field obtained by fitting, and we contend that the magnetic field in the core region may not be generated by the amplification of the magnetic field of the interstellar medium by the shock. Magnetic fields may come from near the center of the black hole or from the accretion disk (\citealt{2006ARA&A..44..463H}).
 The magnetic field strength in the inner accretion disk region can be estimated by comparing it with the magnetic field strength of the BH binary (\citealt{2000Sci...287.1239Z} found $B_{\mathrm{AGN}}\sim10^{-4}B_{\mathrm{Binary}}\sim10^4\mathrm{G}$). This magnetic field is much stronger than the value found in the jet, so the magnetic field in the jet appears to be carried by accretion flow, but is significantly diluted as the jet spreads out and expands (\citealt{2012ApJ...752..157Z}).

\begin{figure}[ht!]
	\centering
	\includegraphics[width=0.6\linewidth]{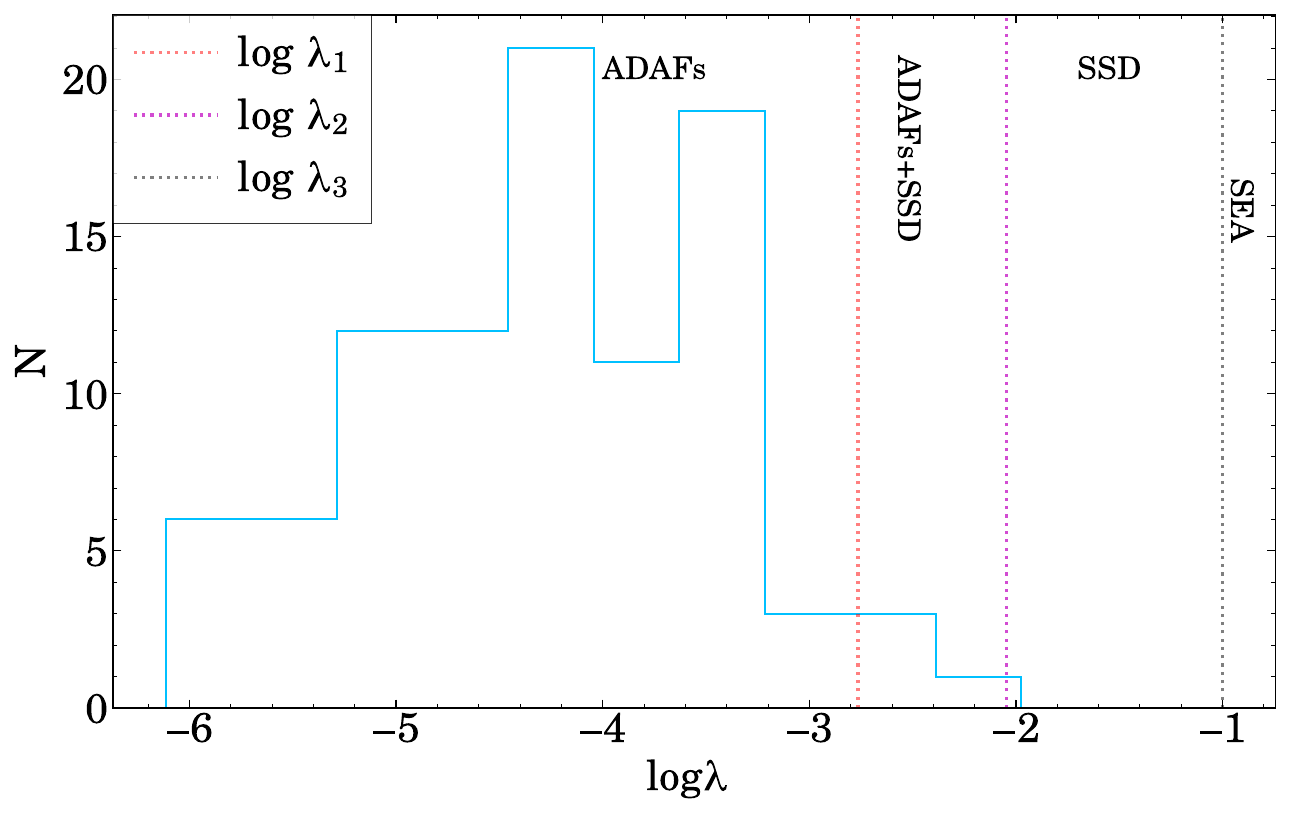}
	\caption{The distribution of $\lambda$ for 94 HBLs. The red dotted line is $\lambda_1$, the purple dotted line is $\lambda_{2}$, and the gray dotted line is $\lambda_{3}$. There are four areas in the picture: $\lambda<\lambda_{1}$ is pure ADAFs, $\lambda_{1}\leq\lambda\leq\lambda_{2}$ is ADAFs+SSD, $\lambda_{2}\leq\lambda\leq\lambda_{3}$ is SSD, and $\lambda\geq\lambda_{3}$ is SEA.}
	\label{fig:6}
\end{figure}

\begin{figure*}[!htbp]
	\centering
	\subfigbottomskip=2pt
	\subfigcapskip=-5pt
	\begin{adjustwidth}{-0.3cm}{-1cm}
	\subfigure{
		\includegraphics[width=0.46\linewidth]{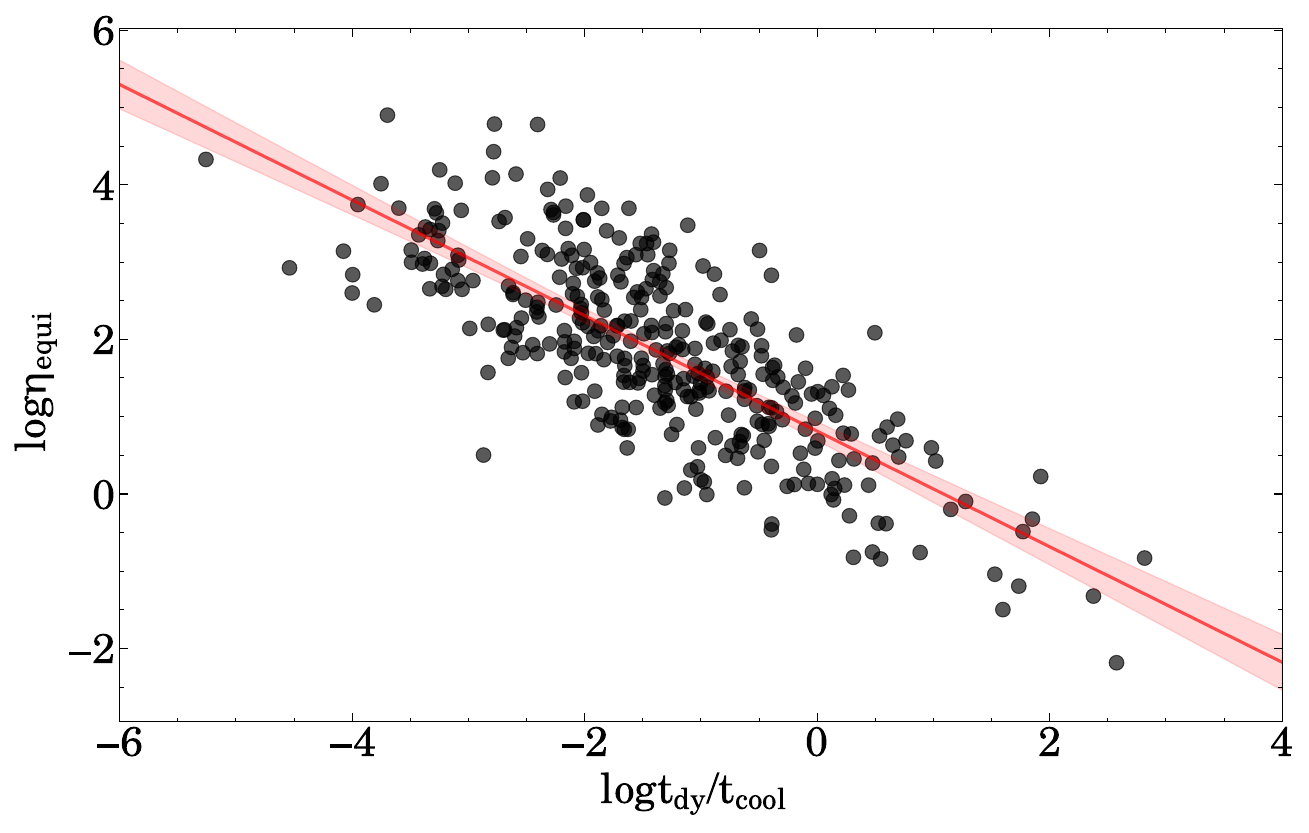}}
	\subfigure{
		\includegraphics[width=0.46\linewidth]{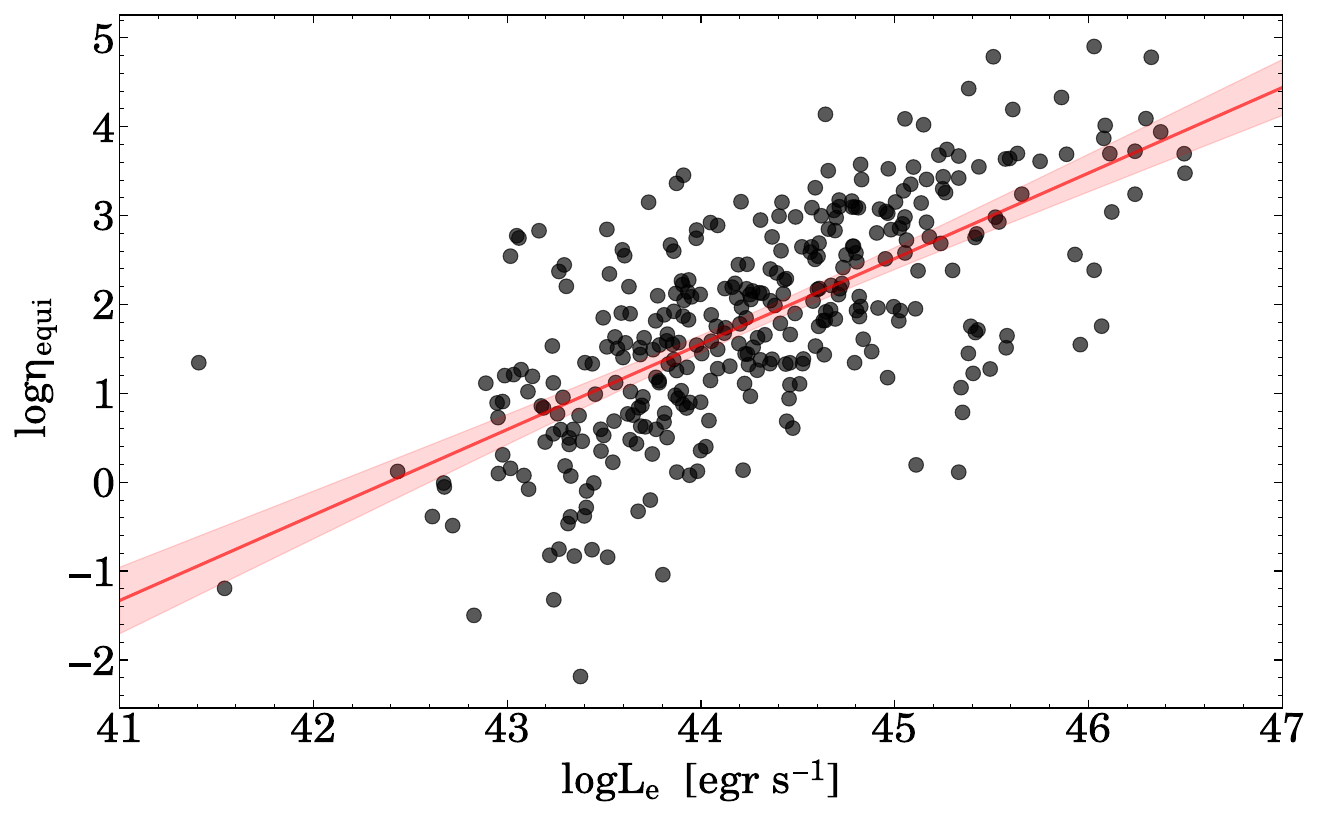}}
	\end{adjustwidth}
	
\caption{Left panel: the relationship between the equipartition coefficient (y-axis) and the ratio of dynamic time scale to cooling time scale
		 (x-axis): $\log\eta_{\mathrm{equi}}=(-0.75\pm0.03)\log t_{\mathrm{dyn}}/t_{\mathrm{cool}}+(0.81\pm0.06)$.
Right panel: the relation between the equipartition coefficient and  the jet power of relativistic electrons: $\log\eta_{\mathrm{equi}}=(0.96\pm0.06)\log L_{\mathrm{e}}-(40.76\pm2.49)$.}
	\label{fig:7}
\end{figure*}

\begin{figure*}[!htbp]
	\centering
	\subfigbottomskip=2pt
	\subfigcapskip=-5pt
	\begin{adjustwidth}{-0.0cm}{1cm}
		\subfigure{
			\includegraphics[width=0.34\linewidth]{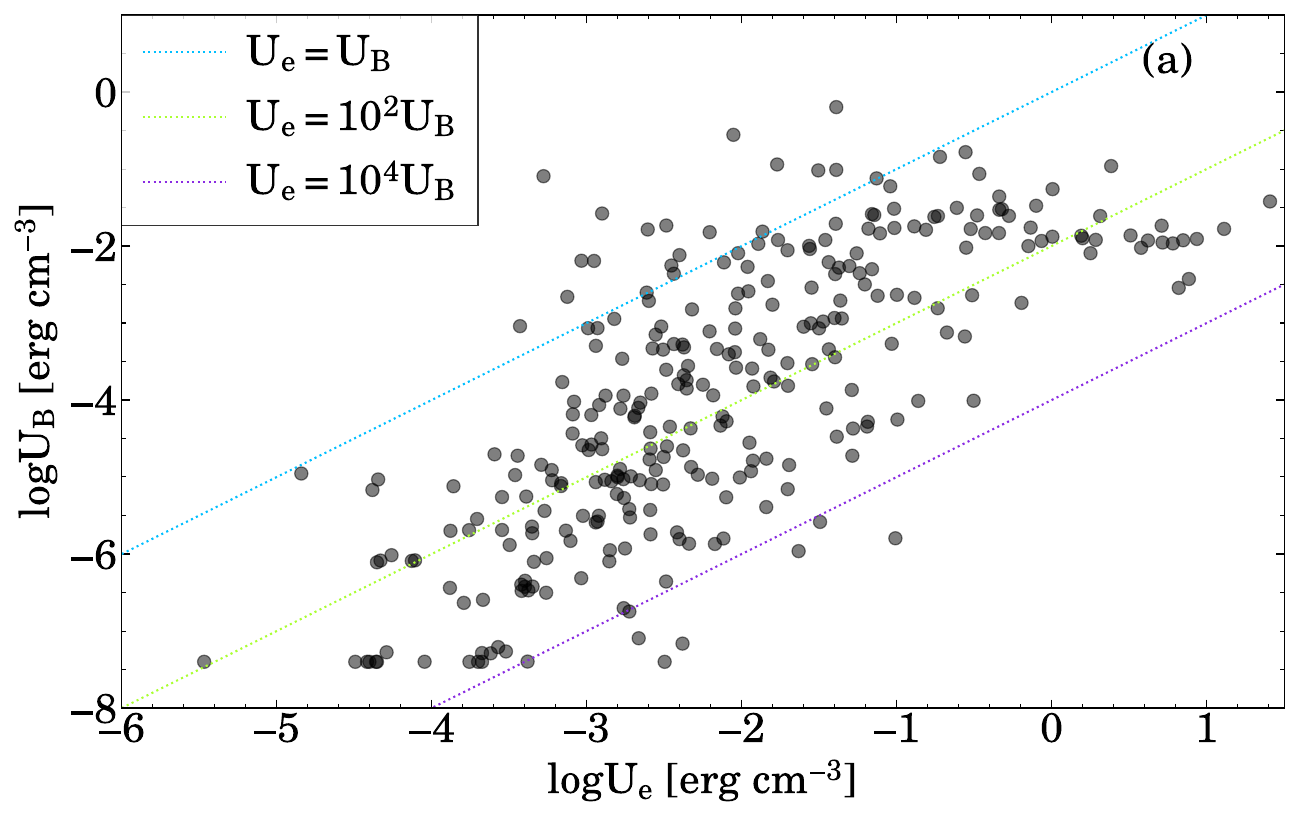}}\hspace{-1mm}
		\subfigure{
			\includegraphics[width=0.34\linewidth]{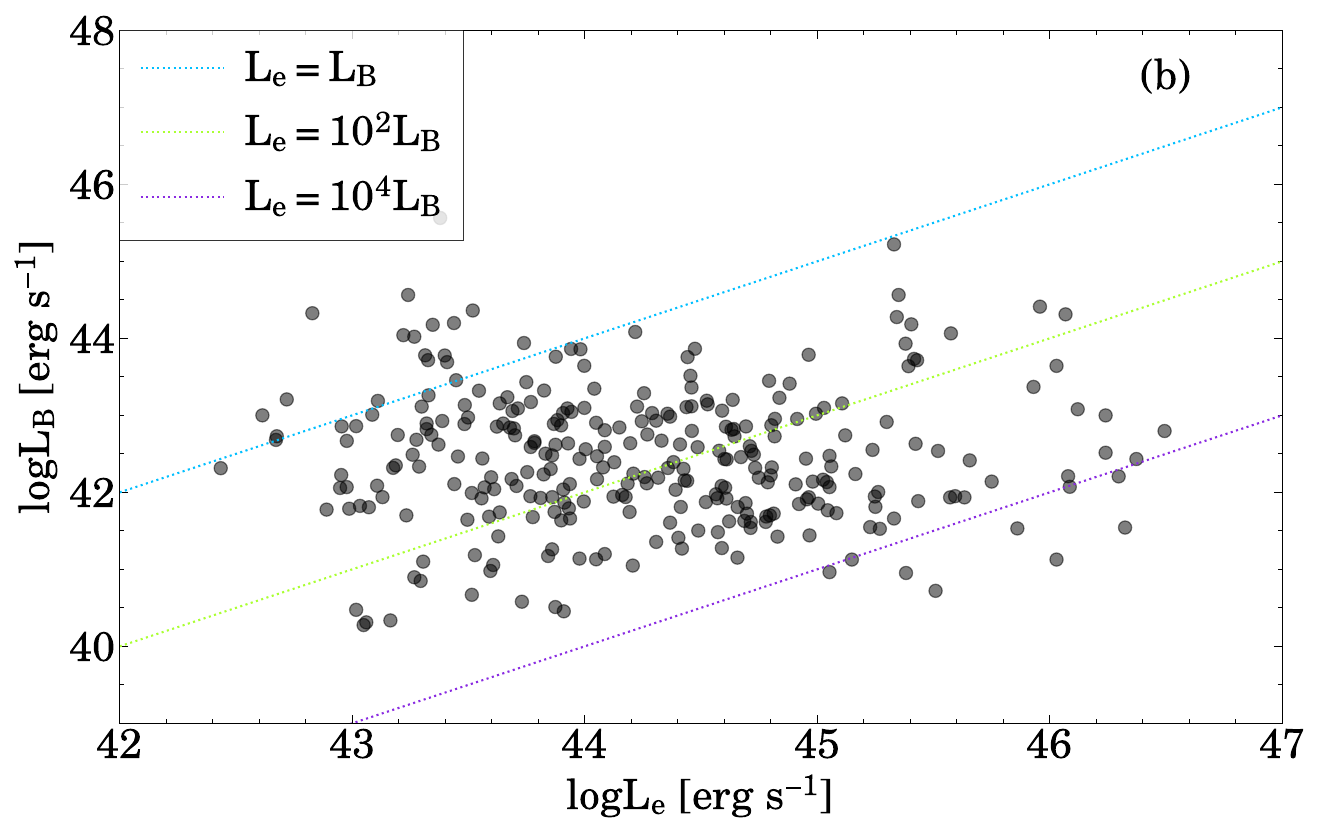}}\hspace{-2mm}
		\subfigure{
			\includegraphics[width=0.34\linewidth]{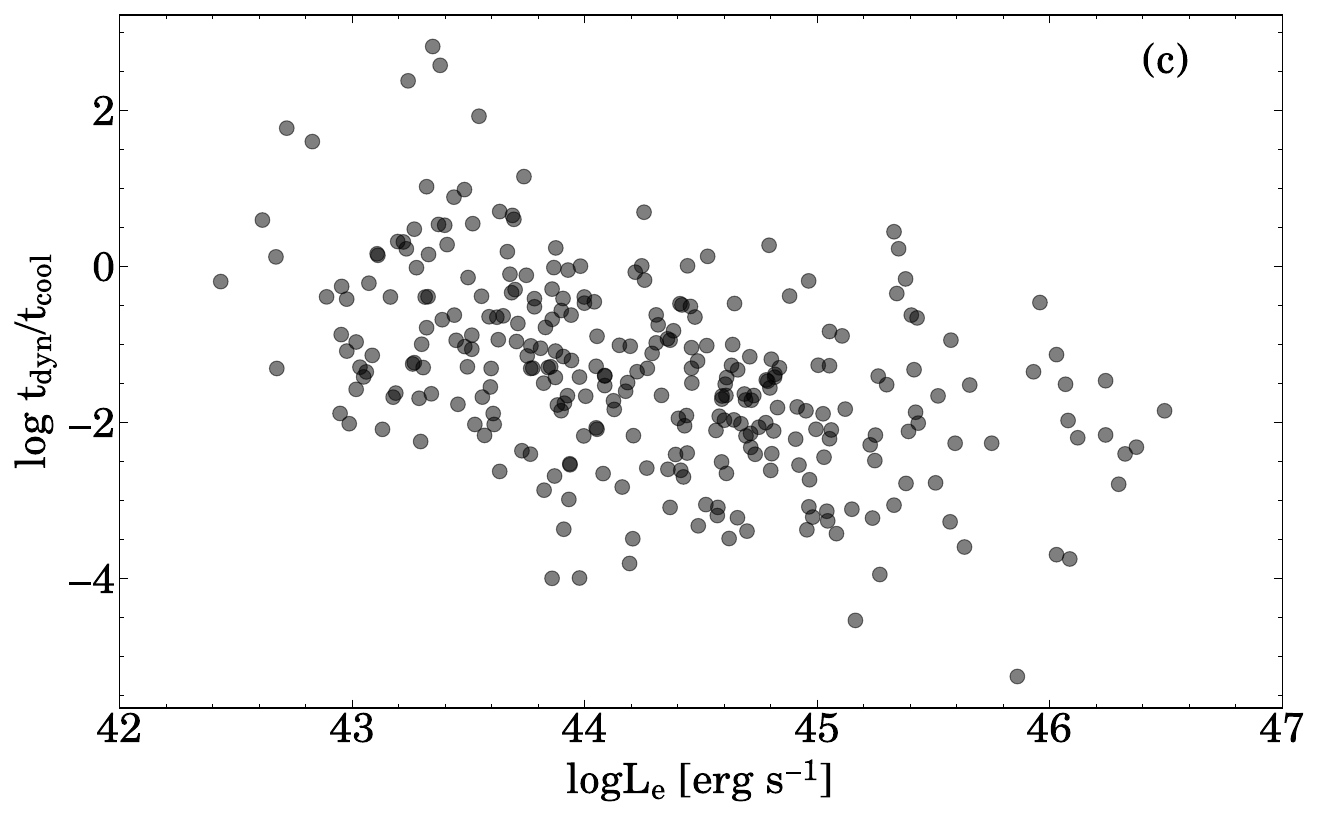}}
	\end{adjustwidth}
	
\caption{(a) Correlation between electron energy density and magnetic energy density across 299 sources with available redshift.
(b) Correlation between jet power of electrons and jet power of magnetic field  across 299 sources with available redshift.
(c) Correlation between the ratio of dynamic time scale to cooling time scale and jet  power of electrons across 299 sources with available redshift.}
	\label{fig:8}
\end{figure*}

Similarly, Figure~\ref{fig:5} on the left compares the relativistic electron power $L_e$ and magnetic field power $L_B$.
We observe that the vast majority of HBLs are located in the $L_e$$>$$L_B$ region, and we also consider the radiation efficiency of the jet.
In the right  of Figure~\ref{fig:5}, we give the ratio of the dynamic time scale $t_{dyn}=R/c$ to the electron cooling time scale $t_{cool}$ as a function of $L_e$.
In our model, we choose the smaller of the synchrotron radiation cooling scale and the SSC cooling scale.
The expression is as follows (\citealt{1986rpa..book.....R,2019MNRAS.484.1192S}):
\begin{equation}t_\mathrm{cool}^{Syn}=\frac{\gamma_\mathrm{b}m_\mathrm{e}c^2}{P_\mathrm{Syn}}\end{equation}
\begin{equation}t_\mathrm{cool}^{SSC}=\frac{\gamma_\mathrm{b}m_\mathrm{e}c^2}{P_\mathrm{SSC}}\end{equation}
where \begin{equation}P_{\mathrm{Syn}}=\frac{4}{3}\sigma_{\mathrm{T}}cU_{\mathrm{B}}\gamma_{\mathrm{b}}^{2},\end{equation}
\begin{equation}P_{\mathrm{SSC}}=\frac{4}{3}\sigma_{\mathrm{T}}cU_{\mathrm{\gamma}}\gamma_{\mathrm{b}}^{2},\end{equation}
where $U_{\gamma}=L_{\mathrm{s}}/4\pi R^{2}c\delta^{4}$ is the radiation energy density of the synchrotron photons and $L_{\mathrm{s}}=(4\pi R^3/3)\delta^{4}\int N\left(\gamma\right)P_{\mathrm{Syn}}\mathrm{d}\gamma $ is the synchrotron peak luminosity. So,
\begin{equation}t_\mathrm{cool}=\min\left[t_{\mathrm{cool}}^{Syn}, t_{\mathrm{cool}}^{SSC}\right]\end{equation}
 Most HBLs ($86.5\%$) are located in the $t_{cool}$$<$$t_{dyn}$ region, which indicates that the jet radiation efficiency of HBLs is inferior. Therefore, we argue that HBLs may have optically thin advection-dominated accretion flows (ADAFs;   
\citealt{1994ApJ...428L..13N,2002ApJ...570L..13C,2003ApJ...599..147C,2003A&A...409..887W,2023MNRAS.526.4079C}).
 
Moreover, we also discuss the accretion rates of a fraction of the HBLs for which we collected black hole masses and accretion disk luminosity from \cite{2021ApJS..253...46P}. However, only 94 HBLs in our sample have black hole masses and accretion disk luminosity. This is shown in Table \ref{tab:4}.

Through the `line accretion rate' we can determine the accretion disk condition of the HBL. \cite{2002ApJ...579..554W} define the formula for the `line accretion rate' $\lambda$ as follows:
 \begin{equation}\lambda=\frac{L_{\mathrm{lines}}}{L_{\mathrm{Edd}}}\end{equation},
 where $L_{\mathrm{lines}}=\xi L_{\mathrm{disk}}$. Assuming that most of the line luminosity ($L_{\mathrm{lines}}$ ) is photoionized by the accretion disk (\citealt{Netzer1990}), the $L_{\mathrm{lines}}$ should be proportional to the total luminosity of the accretion disks. \cite{Netzer1990} defined $\xi \sim 0.1$. $L_{\mathrm{Edd}}$ is the Eddington luminosity, $L_{\mathrm{Edd}}=1.3\times10^{38}(M_{\mathrm{BH}}/M_{\odot})~\mathrm{erg~s}^{-{1}}$. \cite{2002ApJ...579..554W} found the relation between $\lambda$ and the dimensionless accretion rate ($\dot{m}$) for an optically thin ADAF as follows:
 \begin{equation}\dot{m}=2.17\times10^{-2}\alpha_{0.3}\xi_{-1}^{-1/2}\lambda_{-4}^{1/2},\end{equation},
 where $\alpha_{0.3}=\alpha/0.3,~\xi_{-1}=\xi/0.1$, and $ \lambda_{-4}=\lambda/10^{-4}$ (\citealt{2003A&A...409..887W}). The viscosity parameter $\alpha$ is 0.3 (\citealt{1995ApJ...452..710N}). \cite{1998tbha.conf..148N} suggested that an optically thin ADAF appears when $\dot{m}\leq\alpha^{2}$. So, Equation (12) can then be rewritten as
 \begin{equation}\lambda_1=1.72\times10^{-3}\xi_{-1}\alpha_{0.3}^2.\end{equation}
 Optically thin ADAFs require $\lambda<\lambda_1$.
 When $1>\dot{m}\geq\alpha^{2}$, the disk has a standard optically thick, geometrically thin structure (\citealt{1973A&A....24..337S}). \cite{2003A&A...409..887W} has,
 \begin{equation}\dot{m}=\frac{L_{\mathrm{line}}}{\xi L_{\mathrm{Edd}}}=10\xi_{-1}^{-1}\lambda,\end{equation}
 and
 \begin{equation}\lambda_2=9.0\times10^{-3}\xi_{-1}\alpha_{0.3}^2,\end{equation}.
 When $\lambda\geq\lambda_{2}$, the standard disk (SSD) can exist. Furthermore, when $\lambda$ is between $\lambda_1$ and $\lambda_2$, the accretion flow may be in a state where the standard disk coexists with the ADAF (\citealt{2003A&A...409..887W}) The possibility of a mixed state of AGN accretion disks has already been discussed (\citealt{1999ApJ...525L..89Q,2000ApJ...541..120H,2000A&A...360.1170R}), and \cite{2000astro.ph..8319G}, \cite{2024ApJ...962..122K}, and \cite{2024arXiv240217099R} suggest that a transition from SSD to the ADAF is possible, perhaps in the form of evaporation (\citealt{1999ApJ...527L..17L}). The transition radius depends on the accretion rate, black hole mass, and viscosity. However, the structure of the disk in such a state is complex, mainly due to uncertainties in the viscosity. If $\dot{m}\geq1$, we can get
 \begin{equation}\lambda_3=0.1\xi_{-1},\end{equation}.
 When $\lambda\geq\lambda_{3}$, the slim disk can exist, and so-called super-Eddington accretion flow (SEA).
 
Based on the distribution of $\lambda$, we can determine the condition of the HBL's accretion disk. As shown in Fig \ref{fig:6}, 91 ($96.8\%$) of our 94 HBLs may be purely optically thin ADAFs, 2 ($2.1\%$) belong to the mixed state of ADAFs+SSD, and the rest are SSD. Our results are consistent with \citep{2002ApJ...579..554W,2003A&A...409..887W} and \cite{2023MNRAS.526.4079C}. To sum up, our results suggest HBLs may have optically thin ADAFs.

 In the left panel of Figure \ref{fig:7}, we present the equipartition coefficient of HBLs and illustrate its relationship to the ratio of the dynamic timescale to cooling timescale.
 We have observed a significant correlation between these two factors across all HBLs  ($r=-0.75$, $P=2.16\times10^{-69}$).
 Specifically, HBLs with high radiative efficiency tend to exhibit an equipartition coefficient closer to unity.
 Conversely, our findings indicate a notable positive correlation between the equipartition coefficient and the relativistic electron carrying power ($r=0.68$, $P=4.26\times10^{-48}$).
 As shown in the right panel of Figure \ref{fig:7}, HBLs with lower relativistic electron power display correspondingly smaller values for their equipartition coefficients.

\cite{2016MNRAS.456.2374T} studied the electron and magnetic field energy densities of 45 BL Lac objects using the same  one-zone lepton model, and they found that most of the sources are characterized by an electron component strongly dominating over the magnetic one, with an average ratio $U_{\mathrm{e}}/U_{B}\sim100$ (see Figures 3 and 4 of \citealt{2016MNRAS.456.2374T}).
We have expanded the samples and validated their results.

In addition, 50 sources in our sample do not have available redshifts.
As shown  Figure \ref{fig:8}, we perform a similar analysis excluding these sources.
Clearly, even excluding these sources did not make much difference to our results and is consistent with previous studies.

\subsection{Energy budget}  \label{sec:4.3}

\begin{figure}[htbp]
	\centering
	\subfigbottomskip=2pt
	\subfigcapskip=-5pt
	
	\begin{adjustwidth}{-0.3cm}{-1cm}
		\subfigure{
			\includegraphics[width=0.46\linewidth]{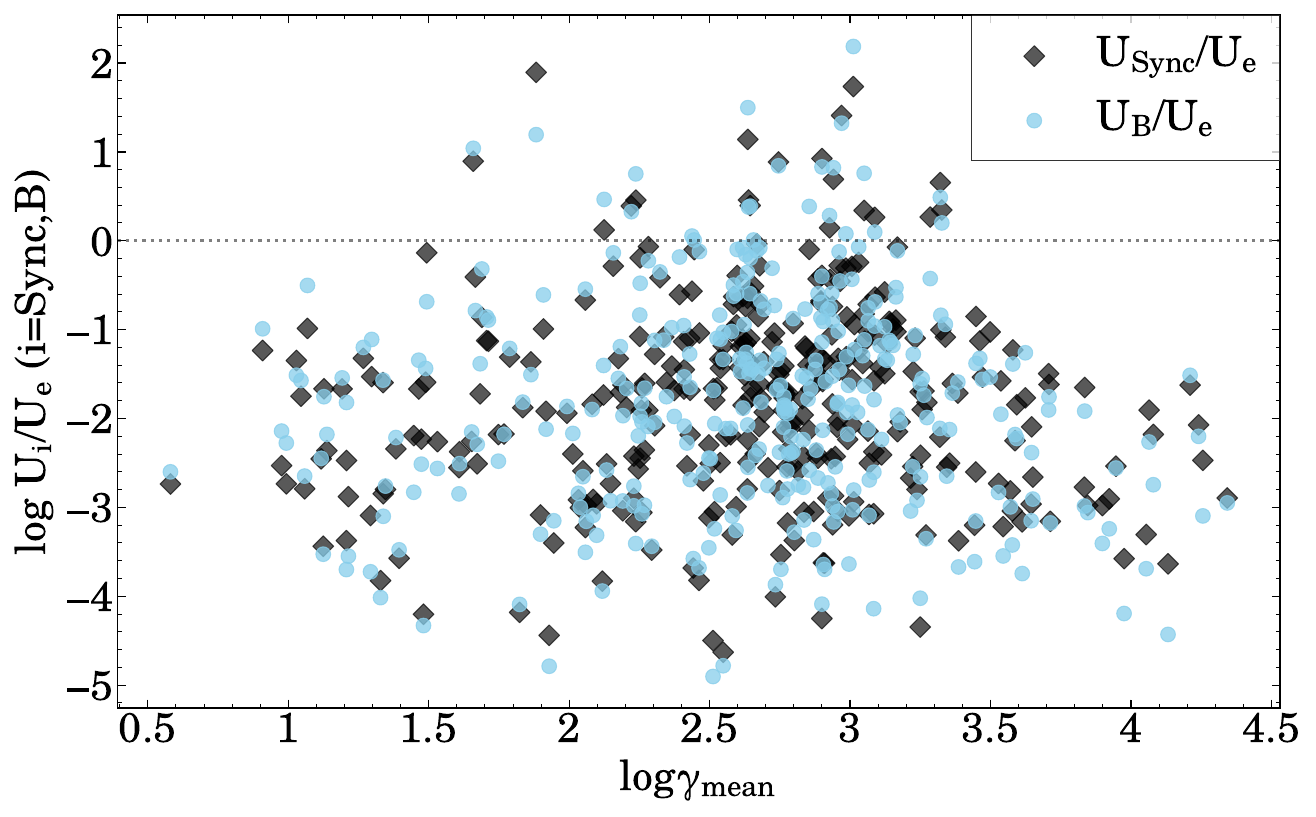}}
		\subfigure{
			\includegraphics[width=0.46\linewidth]{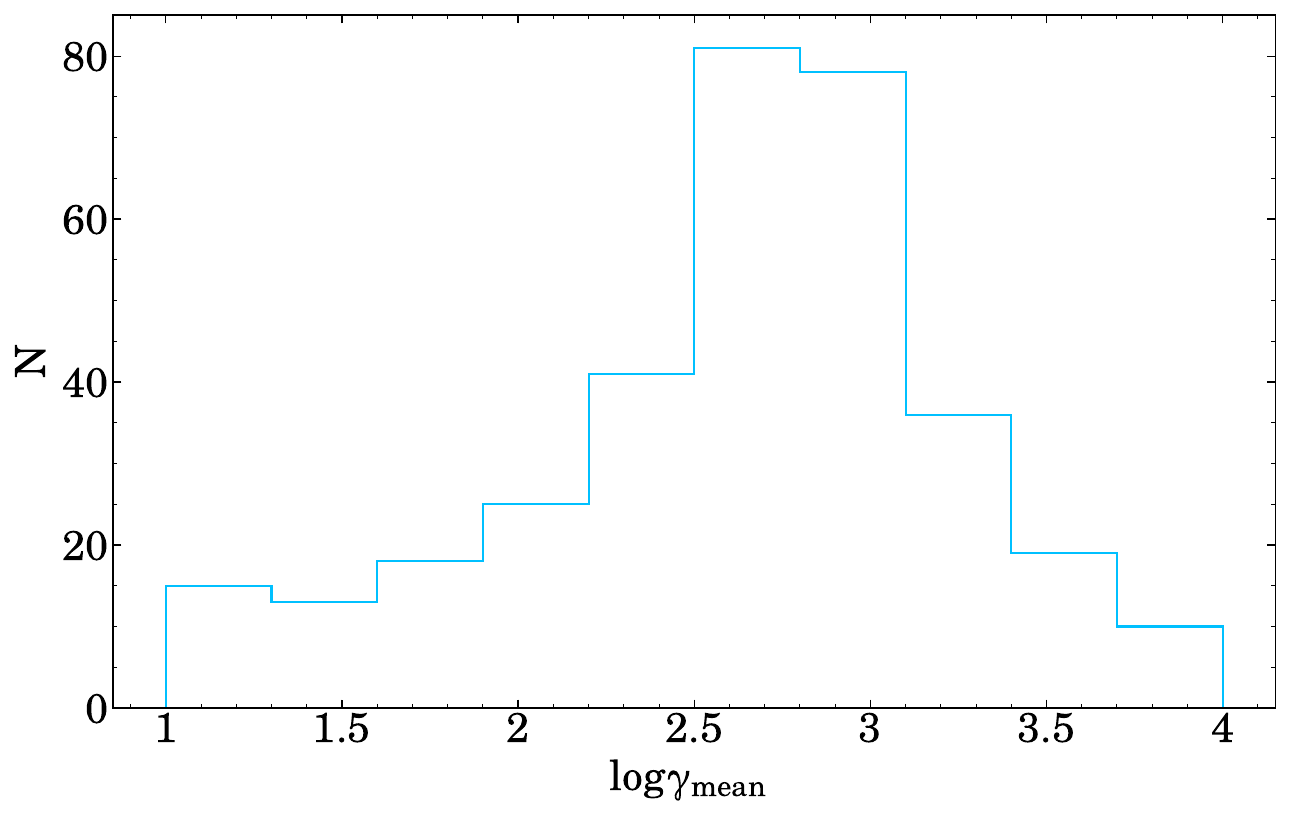}}
	\end{adjustwidth}
	
	\caption{Lrft panel: the $U_i/U_e (i=Sync,B)$ as a function of $\gamma_{mean}$.
		Right panel: the distribution of the average electrons Lorentz factor $\gamma_{mean}$.}
	\label{fig:9}
\end{figure}
\cite{2018MNRAS.478.3855Z} discussed Mrk 501, a typical HBL, for which they plot the predicted comoving energy density of ultrarelativistic electrons (the Figure 3).
By comparison, they discovered that most of the energy is stored in low-energy electrons, and the average Lorentz factor of electrons is $\gamma_{mean}=2447$.   

\begin{figure*}[htbp]
	\centering
	\subfigbottomskip=2pt
	\subfigcapskip=-5pt
	\begin{adjustwidth}{-0.0cm}{1cm}
		\subfigure{
			\includegraphics[width=0.35\linewidth]{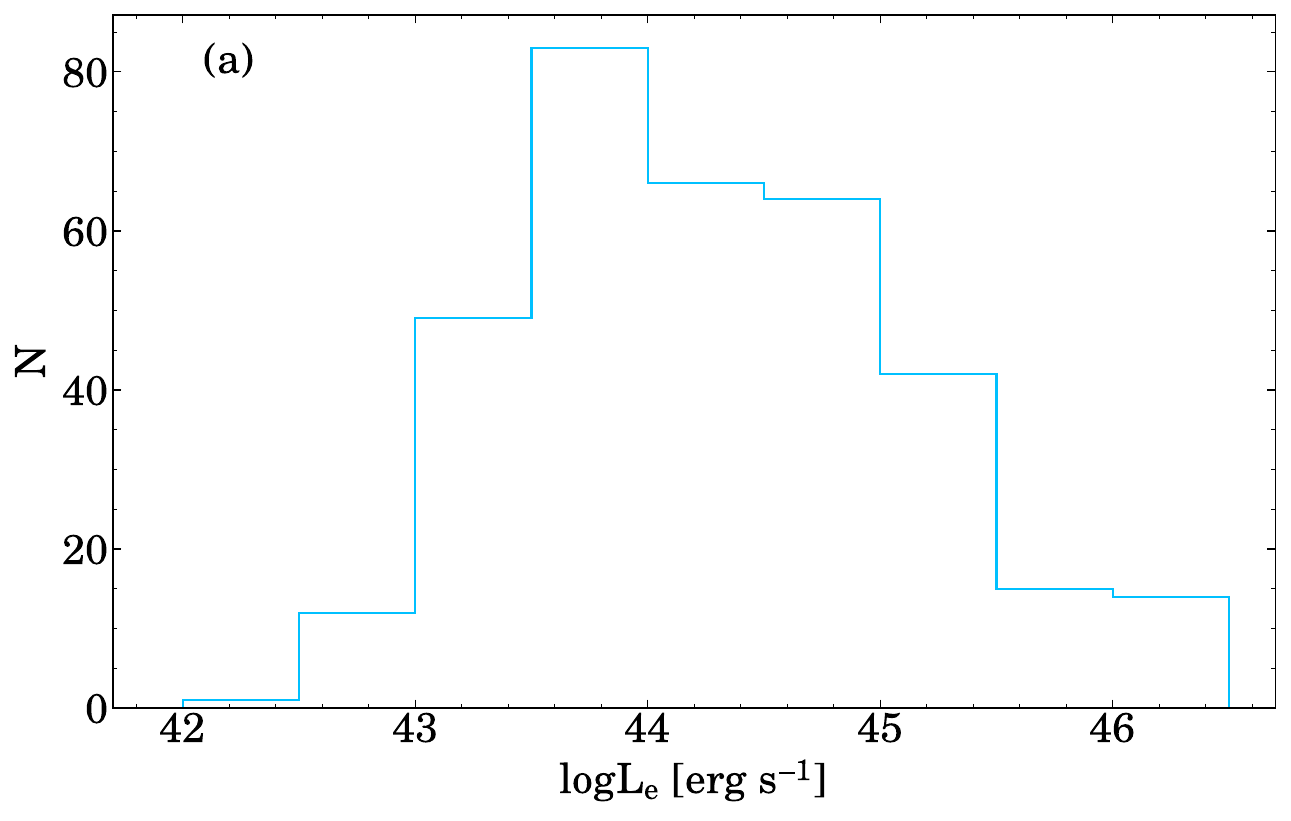}}\hspace{-1mm}
		\subfigure{
			\includegraphics[width=0.35\linewidth]{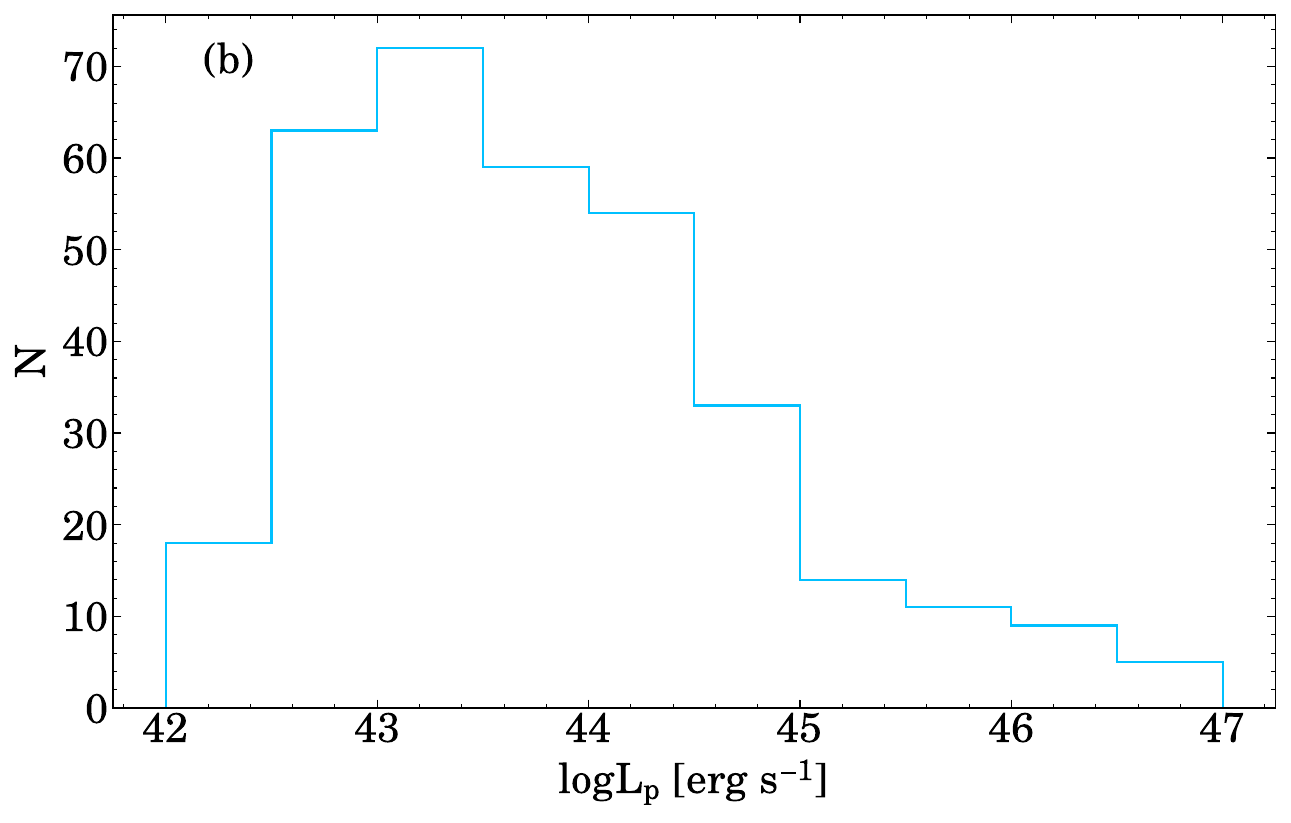}}\hspace{-1mm}
		\subfigure{
			\includegraphics[width=0.35\linewidth]{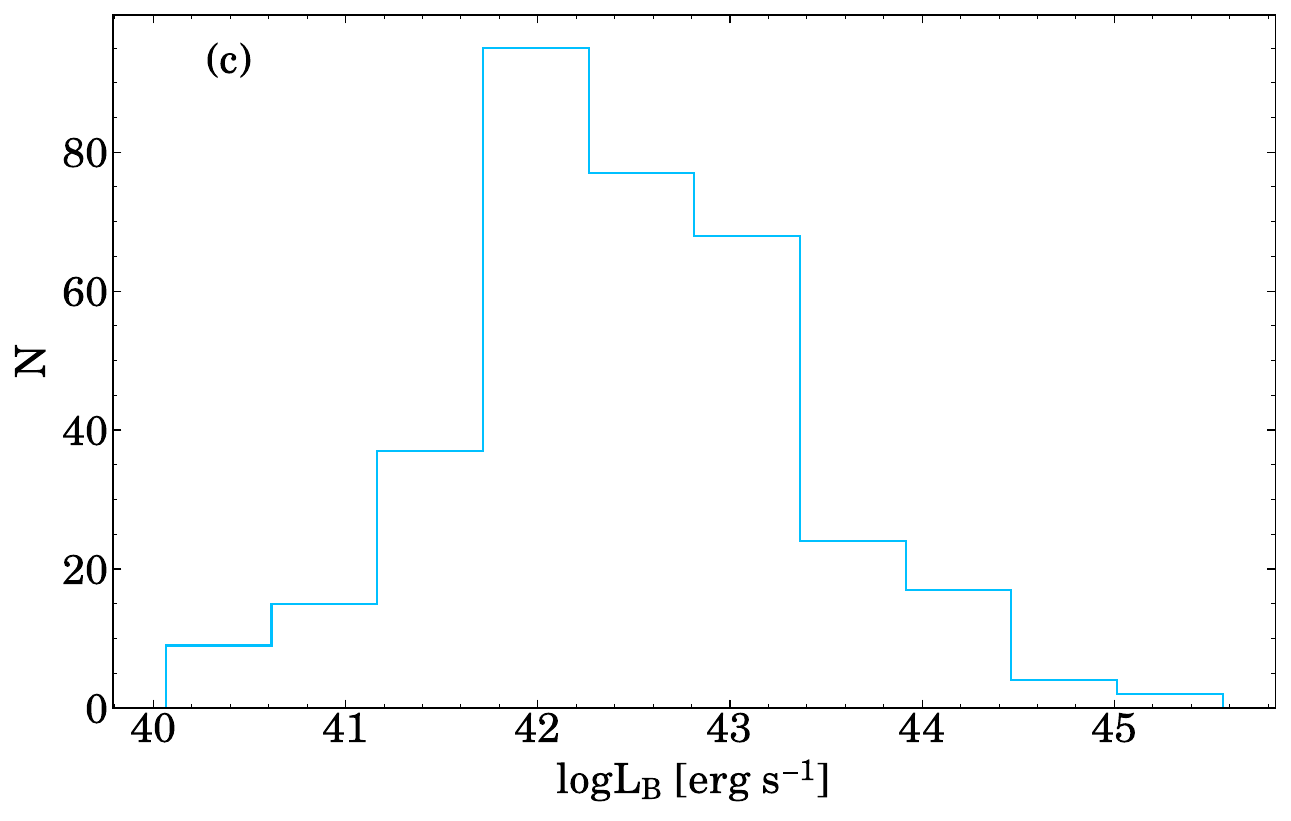}}\hspace{-1mm}
		\subfigure{
			\includegraphics[width=0.35\linewidth]{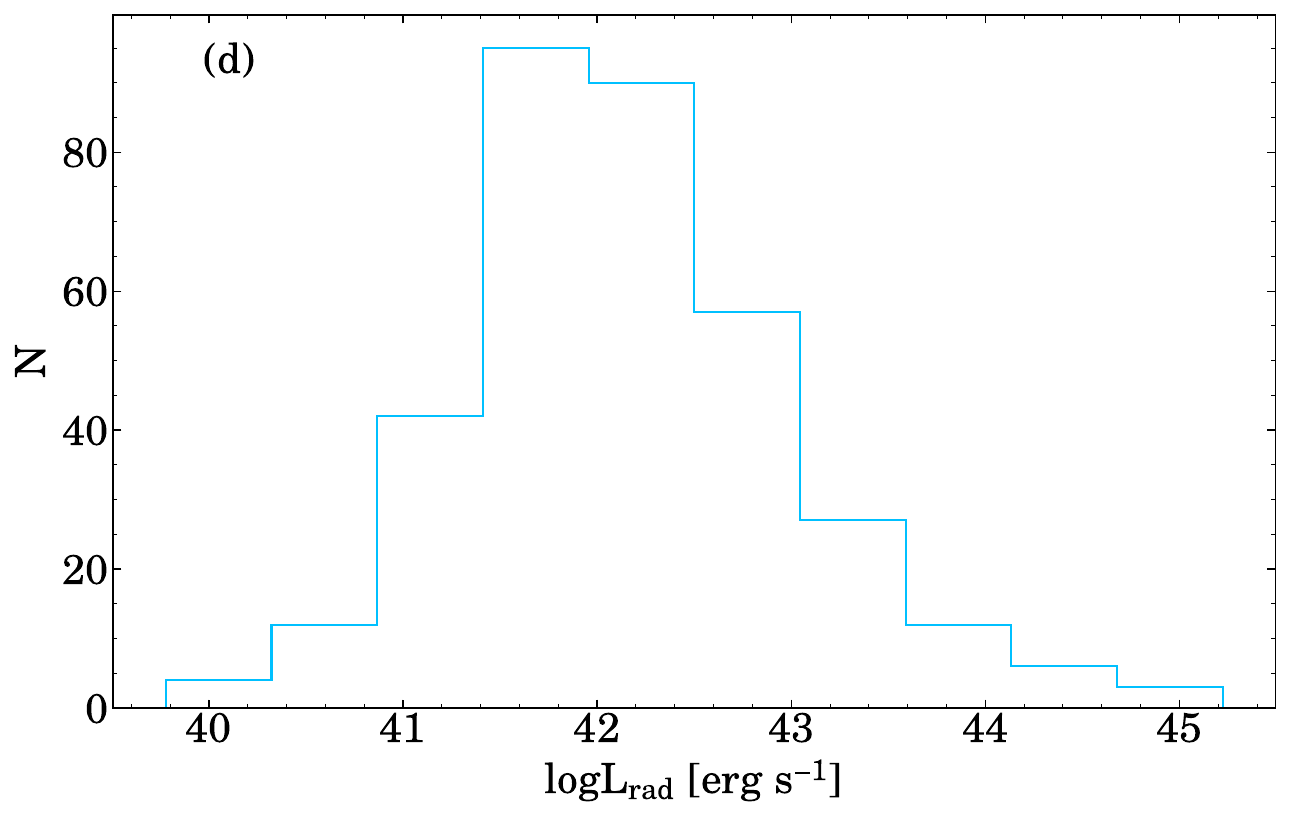}}\hspace{-0.9mm}
		\subfigure{
			\includegraphics[width=0.35\linewidth]{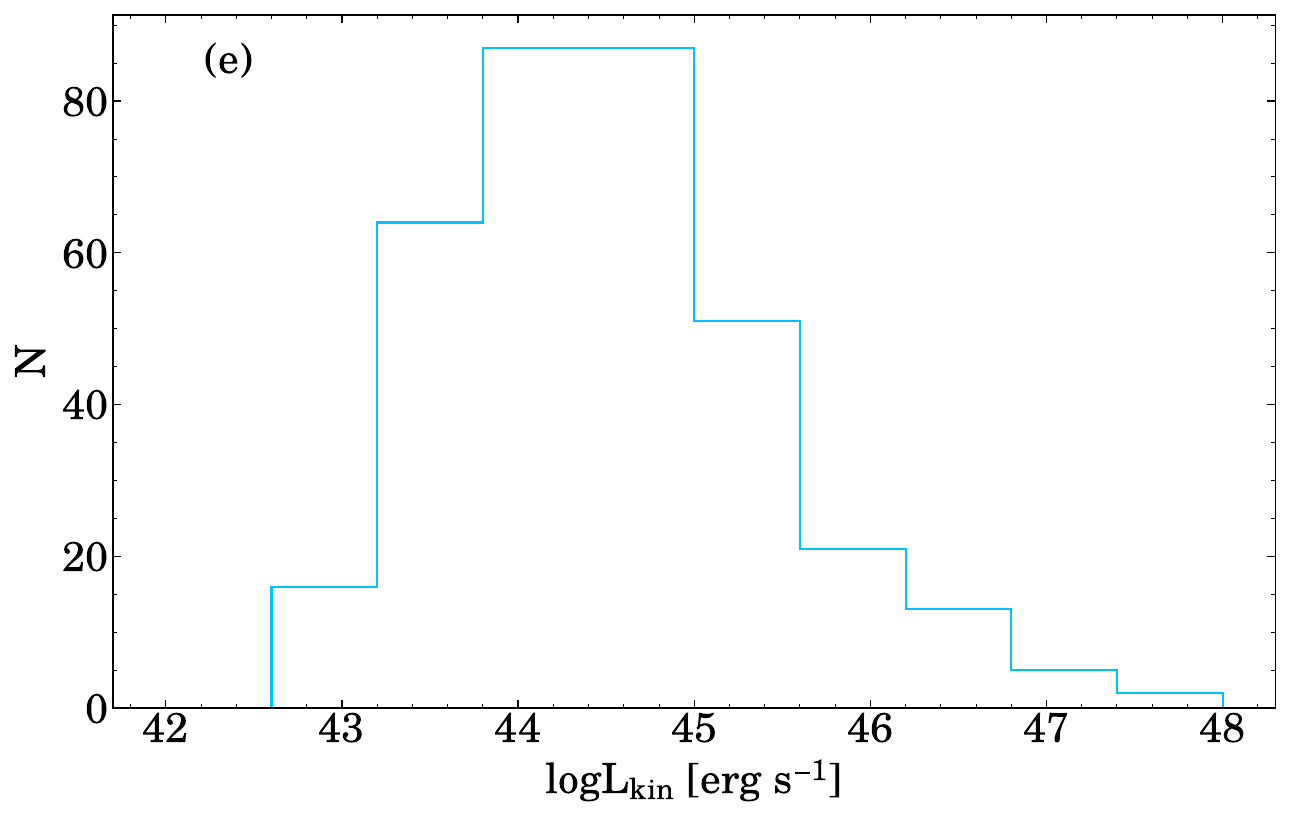}}
	\end{adjustwidth}
	\caption{Distribution of the jet power of (a) relativistic electrons, (b) cold protons, (c) magnetic fields, (d) radiation, and (e) the kinetic power.}
	\label{fig:10}
\end{figure*}

In this paper, we plot the ratio of magnetic field energy density and synchrotron radiation energy density to electron energy density as a function of the mean electron Lorentz factor in the left panel of Figure~\ref{fig:9} and the distribution of the mean electron Lorentz factor $\gamma_{\operatorname{mean}}\simeq\gamma_{\min}\ln\left(\gamma_{\mathrm{b}}/\gamma_{\min}\right)$ in the right panel.
We found that:
(1) most  HBLs have $U_{e}>U_{Syn}\sim U_{B}$.
(2) $\log\gamma_{mean}$  is in the range  $(1\sim4)$, and the mean value is $\langle\log\gamma_{mean}\rangle\simeq2.75$.
Our results are consistent with \cite{2018MNRAS.478.3855Z} in that most of the energy of the HBLs is stored in low-energy electrons.

\begin{sidewaystable*}[!htp]
	\tablenum{5}
	\centering
	\small
	\begin{center}
		\caption{Energy Density and jet power 	\label{tab:5} }    
		\setcounter{table}{5}
		\renewcommand{\thetable}{5/arabic{table}}
		\renewcommand\arraystretch{2}
		\begin{adjustwidth}{-2cm}{0cm}
			\scalebox{1.3}{
				\begin{tabular}{cccccccccc}
					\hline\hline	 			
					$\rm 4FGL\;Name$  & $ log U_e$ &  $ log U_B$ & $ log U_p$ & $ log U_{Sync}$ &  $ log L_e$ &  $ log L_B$  &  $ log L_p$  &  $ log L_{Sync}$  & { $ log L_{SSC}$} \\
					\normalsize(1) & \normalsize(2) & \normalsize(3) &\normalsize(4) & \normalsize(5)   &\normalsize(6) &\normalsize(7) &\normalsize(8) &\normalsize(9) &\normalsize(10)  \\
					\hline
					\centering
					{$ \rm J0013.9-1854$} & -1.13  & -1.12  & -1.35  & -1.64  & 42.67  & 42.68  & 42.45  & 41.69  & 40.68  \\
					{$ \rm J0014.1-5022$} & 0.38  & -0.96  & -0.05  & -0.87  & 41.41  & 40.06  & 40.97  & 39.67  & 39.11  \\
					{$ \rm J0022.0+0006$} & -0.72  & -0.84  & -0.68  & -1.76  & 42.43  & 42.31  & 42.47  & 40.92  & 39.60  \\
					{$ \rm J0030.2-1647$} & -3.35  & -6.42  & -3.48  & -6.41  & 44.92  & 41.85  & 44.79  & 41.39  & 41.05  \\
					{$ \rm J0033.5-1921$} & -3.13  & -5.70  & -2.24  & -5.40  & 45.93  & 43.37  & 46.83  & 43.19  & 43.09  \\
					{$ \rm J0043.7-1116$} & -2.38  & -4.65  & -2.51  & -4.91  & 44.43  & 42.15  & 44.30  & 41.42  & 40.84  \\
					{$ \rm J0045.3+2128$} & -3.29  & -4.84  & -3.05  & -4.62  & 45.96  & 44.41  & 46.20  & 44.15  & 43.91  \\
					{$ \rm J0051.2-6242$} & -4.35  & -5.03  & -4.51  & -4.48  & 44.44  & 43.76  & 44.27  & 43.84  & 43.69  \\
					{$ \rm J0054.7-2455$} & -2.10  & -4.28  & -1.10  & -4.46  & 44.12  & 41.94  & 45.12  & 41.28  & 40.77  \\
					{$ \rm J0058.3+1723$} & -3.88  & -6.44  & -5.38  & -6.42  & 44.75  & 42.19  & 43.25  & 41.73  & 41.00  \\
					\hline
			\end{tabular}}\\
		\end{adjustwidth}
	\end{center} 
	\tablecomments{Columns (1): the 4FGL name of sources; columns (2-5): the energy density of relativistic electrons, magnetic fields, cold protons, and synchrotron radiation (erg $cm^{-3}$); columns (6)-(10): the power of relativistic electrons, magnetic fields, protons, synchrotron radiation and synchrotron self-Compton (erg $s^{-1}$).
		 (This table is available in its entirety in machine-readable form.)   }		
	
\end{sidewaystable*}

\subsection{Physical Properties of the HBL Jet}

In \textit{JetSet}, we can easily estimate the jet power by fitting SEDs.
The jet kinetic power is carried by electrons, magnetic fields, and cold protons (\citealt{1977MNRAS.179..433B,2014Natur.515..376G,2019MNRAS.489.5076S}), i.e.,

\begin{equation}
	\setlength\abovedisplayskip{16pt}
	\setlength\belowdisplayskip{6pt}
	L_{\mathrm{i}}=\pi R^2\Gamma^2cU_{i}
\end{equation}
\begin{equation}
	\setlength\abovedisplayskip{-6pt}
	\setlength\belowdisplayskip{-6pt}
	L_{\mathrm{kin}}=(L_{e}+L_{B}+L_{p})
\end{equation}
where $U_{i}(i=e,p,B)$ is the energy density.
The total emitted radiative power is $L_{\mathrm{rad}}\simeq(L_{\mathrm{Syn}}+L_{\mathrm{SSC}})$, where $L_{\mathrm{Syn}}$ and $L_{\mathrm{SSC}}$ are the radiative powers of Syn  and SSC respectively (\citealt{2018MNRAS.478.3855Z}). Table~\ref{tab:5} lists our results.
\begin{figure}[ht!]
	\centering
	\includegraphics[width=0.6\linewidth]{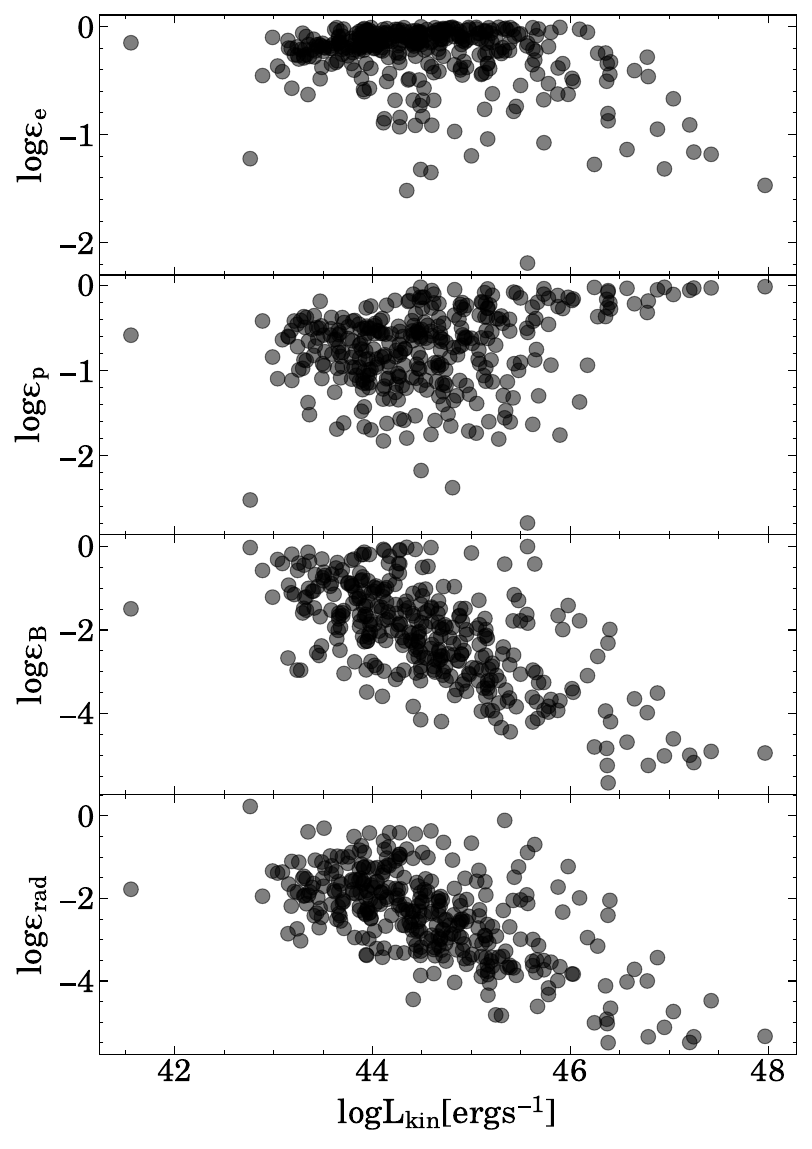}
	\caption{A portion of the total jet power is converted into relativistic electrons, cold protons, and radiation.}
	\label{fig:11}
\end{figure}

Figure~\ref{fig:10} (a)-(e) plot the distribution of $L_{e}$, $L_{B}$, $L_{p}$, $L_{rad}$, and $L_{kin}$.
The range of $L_{e}$ and $L_{p}$ is $10^{42\sim48} erg s^{-1}$.
The range of $L_{B}$ and $L_{rad}$ is $10^{40\sim46} erg s^{-1}$.
We found that $L_{e}\sim L_{p}>L_{B}\sim L_{rad}$.
Our results are consistent with \cite{2023ApJS..268....6C}.

In Figure~\ref{fig:11}, we plot the ratios of the power of relativistic electrons, magnetic fields, cold protons, and radiation to the jet kinetic power, (i.e. $\epsilon_{i}=\frac{L_{i}}{L_{kin}}, i=e,p,B,rad$).
We find that:
(1) The part of the power carried by the magnetic field is equivalent to that carried by the radiation (\citealt{2012ApJ...752..157Z}).
(2) The $\log\epsilon_{B}$ of HBLs is always less than zero, which indicates that the jet kinetic power of HBLs is not affected by Poynting flux (\citealt{2015MNRAS.451..927Z,2017ApJ...851...33P,2023ApJS..268....6C}). (3) The $\log\epsilon_{rad}$ of HBLs is less than zero, which means that the jet kinetic power of blazars is greater than the radiant power.
There is a very low radiant efficiency, as we obtained in Section~\ref{sec:4.2}.
(4) For most sources, the power in the jet is carried by relativistic electrons, $\log\epsilon_{e}>0.5$, consistent with the results we obtain in Section~\ref{sec:4.3}.
However, when the jet kinetic power  is greater than $10^{46} erg s^{-1}$, cold protons dominate it.

\section{Summary}   \label{sec:conclusion}

We have used the open-source program \textit{JetSet} to fit the SEDs of 348 HBL samples.
Several physical parameters of the jet can be obtained, such as the Doppler factor, magnetic field, jet power, etc.
With the above-mentioned information, we further discussed the energy budget and the energy equiseparation of the jets of HBLs.
The main results are summarized as follows.

(1) The one-zone Syn+SSC model can reasonably reproduce the SEDs of HBLs.

(2) We discover that the electron energy density of  HBLs is much higher than the magnetic field energy density.
We contend that the magnetic field in the core region may not be caused by the amplification of the magnetic field of the interstellar medium by shock waves.
The magnetic field may come from near the center of the black hole or from the accretion disk.

(3) Most the HBLs are located in the $t_{cool}$$<$$t_{dyn}$ region and $\log\epsilon_{em}$ is less than zero, which means that the jet kinetic power of blazars is greater than the jet power of radiation, and they have a very low radiant efficiency.
Therefore, we argue that HBLs may have optically thin ADAFs.
The $\log\epsilon_{B}$  of HBLs is less than zero, which indicates that the jet kinetic power of HBLs is not affected by Poynting flux.

(4)  For most HBLs, we find that $U_{e}>U_{Syn}\sim U_{B}$, $L_{e}\sim L_{p}>L_{B}\sim L_{rad}$ and $\log\epsilon_{e}>0.5$.
We contend that most of the energy in the HBLs is stored in low-energy electrons.
When the jet kinetic power is greater than $10^{46} erg s^{-1}$, it is dominated by cold protons.

\begin{acknowledgments}
We thank the anonymous referee for insightful comments and constructive suggestions. We acknowledge the use of data, analysis tools, and services from the Open Universe platform, the ASI Space Science Data Center (SSDC), the \textit{JetSet} Tools, the Astrophysics Data System (ADS), and the SIMBAD. 
This work is supported by the National Natural Science Foundation of China (grant Nos.12363002, 12163002). 
The authors would like to express their gratitude to EditSprings (https://www.editsprings.cn ) for the expert linguistic services provided.
\end{acknowledgments}

\software{\textit{JetSet} (\citealt{2009A&A...501..879T,2011ApJ...739...66T,2020ascl.soft09001T}), iminuit (V2.22.0) \citep{dembinski_2023_8070217}, Astropy \citep{2013A&A...558A..33A,2018AJ....156..123A,2022ApJ...935..167A}, Matplotlib \citep{2007CSE.....9...90H}, Numpy \citep{2020Natur.585..357H}.
}

\bibliography{sample631}{}
\bibliographystyle{aasjournal}

\end{document}